%% file: main.tex
\documentclass[twoside,10pt]{article}

\usepackage{jmlr2e}
\usepackage{abstract}
\usepackage{color}
\usepackage{array}
\usepackage{bbding}
\usepackage{multirow}
\usepackage[fleqn]{amsmath}
\usepackage{lscape}
\usepackage{comment}
\usepackage{bm}
\usepackage[title]{appendix}
\usepackage{longtable}
\usepackage{mathtools}
\usepackage{dsfont}
\usepackage{fancyhdr}
\usepackage{cleveref}
\usepackage{stackengine}
\usepackage{textcomp}
\usepackage{stfloats}
\usepackage{verbatim}
\usepackage{xcolor}
\usepackage{subfigure}
\usepackage{booktabs} 
\usepackage{bbm}

\usepackage{amsfonts}
\usepackage{xurl}
\usepackage{stackengine}
\usepackage{tikz}
\usetikzlibrary{decorations.pathreplacing}
\usetikzlibrary{positioning,arrows.meta,quotes}
\usetikzlibrary{shapes,snakes}
\usetikzlibrary{bayesnet}
\tikzset{>=latex}
\tikzstyle{plate caption} = [caption, node distance=0, inner sep=0pt, below left=5pt and 0pt of #1.south]
\usepackage[normalem]{ulem}
\usepackage{multirow}

\firstpageno{1}

\begin{document}

\title{Calibrating Car-Following Models via Bayesian Dynamic Regression}

\author{\name Chengyuan Zhang \email enzozcy@gmail.com \\
       \addr Department of Civil Engineering\\
       McGill University\\
       Montreal, QC  H3A 0C3, Canada
       \AND
       \name Wenshuo Wang \email wwsbit@gmail.com \\
       \addr Department of Civil Engineering\\
       McGill University\\
       Montreal, QC  H3A 0C3, Canada
       \AND
       \name Lijun Sun\thanks{Corresponding author.} \email lijun.sun@mcgill.ca \\
       \addr Department of Civil Engineering\\
       McGill University\\
       Montreal, QC  H3A 0C3, Canada}

\editor{}

\maketitle

\begin{abstract}
Car-following behavior modeling is critical for understanding traffic flow dynamics and developing high-fidelity microscopic simulation models. Most existing impulse-response car-following models prioritize computational efficiency and interpretability by using a parsimonious nonlinear function based on immediate preceding state observations. However, this approach disregards historical information, limiting its ability to explain real-world driving data. Consequently, serially correlated residuals are commonly observed when calibrating these models with actual trajectory data, hindering their ability to capture complex and stochastic phenomena. To address this limitation, we propose a dynamic regression framework incorporating time series models, such as autoregressive processes, to capture error dynamics. This statistically rigorous calibration outperforms the simple assumption of independent errors and enables more accurate simulation and prediction by leveraging higher-order historical information. We validate the effectiveness of our framework using HighD and OpenACC data, demonstrating improved probabilistic simulations. In summary, our framework preserves the parsimonious nature of traditional car-following models while offering enhanced probabilistic simulations. The code of this work is available at \url{https://github.com/Chengyuan-Zhang/IDM_Bayesian_Calibration}.
\end{abstract}


\section{Introduction}
Car-following behavior plays a critical role in understanding and predicting traffic flow dynamics. Various car-following models have been developed in the literature, including the optimal speed model (OVM) \citep{bando1995dynamical}, Gipps model \citep{gipps1981behavioural}, and the intelligent driver model (IDM) \citep{treiber2000congested} along with its variants \citep{treiber2003memory, derbel2012modified}. These models typically utilize recent observations of the ego vehicle's speed, relative speed to the leading vehicle, and spacing/gap as inputs to an explicitly predefined nonlinear function. This function computes the acceleration or speed as the driver's decision at the current time step. The parsimonious structure of these models offers computational efficiency, interpretability, and analytical connections with macroscopic relations \citep{treiber2010three}. Consequently, car-following model-based simulations have been widely used to gain insights into complex traffic flow dynamics.

Even though car-following models can reproduce important physical phenomena such as shockwaves, it has been highlighted in many recent studies that the parsimonious structure of some models limits their effectiveness in reproducing real-world driving behaviors with high fidelity \citep{wang2017capturing}. While most of these models overlook high-order historical information, leading to inaccurate predictions, there are notable exceptions. For instance, IDM variants that include aspects of recent history, as discussed in \cite{treiber2003memory}. These developments indicate that incorporating observations from the past $1\sim4$ seconds, as also shown in recent studies \citep{zhang2024bayesian}, can enhance the modeling of car-following decisions. However, challenges arise when calibrating parsimonious car-following models based on single-timestamp observations from real-world trajectories, often leading to temporally correlated errors\footnote{Note that \textit{``errors''} pertain to the true data generating process, whereas \textit{``residuals''} are what is left over (i.e., specific values) after having an estimated model. Assumptions such as normality, homoscedasticity, and independence purely apply to the \textit{errors} of the data generating process, but not the model's \textit{residuals}. Readers should distinguish these two terms in the following.}. The assumption of independent errors in the calibration process introduces bias \citep{hoogendoorn2010calibration}. To address this challenge, \cite{zhang2024bayesian} developed a Bayesian calibration framework by modeling errors using Gaussian processes (GPs). Although this approach offers improved and consistent calibration by incorporating memory effects, it may encounter challenges in error modeling, potentially leading to larger variances in the modeled errors. This can result in simulations that, while more reflective of real-world driving behavior due to the memory component, might also exhibit increased noise, thereby affecting the precision and reliability of the simulation outcomes.

Although these models are valuable in understanding traffic flow dynamics, the limitations suggest a need for a more nuanced approach to address the biased calibration of existing car-following models. In this paper, we utilize a dynamic regression framework to model car-following sequence data, incorporating a generative time series model to capture errors. The proposed framework builds upon the classic car-following model, i.e., IDM, retaining the parsimonious features of traditional car-following models while significantly improving prediction accuracy and enabling probabilistic simulations by incorporating higher-order historical information. Here we emphasize that our framework doesn't change the parsimonious structure of the existing models and thus can be applied to various car-following models. The experiments demonstrate the effectiveness of this framework in achieving enhanced predictions and a more accurate calibration of the car-following models. Our results suggest that the driving actions within the past 10 seconds should be considered when modeling human car-following behaviors. This aligns with the literature \citep{wang2017capturing} where 10-s historical information is testified as the best input.

The contributions of this work are threefold:
\begin{enumerate}
    \item We present a novel calibration method for car-following models based on a dynamic regression framework. This method enhances existing parsimonious car-following models by incorporating higher-order historical information without changing the prevailing models. The inclusion of this flexible form enables unbiased calibration.
    \item The framework integrates autoregressive (AR) processes within time series models to handle errors, representing an advancement from the conventional assumption of independent and identically distributed (\textit{i.i.d.}) errors. This enhancement introduces a statistically rigorous approach, offering improved modeling capabilities.
    \item The data generative processes of our framework offer an efficient probabilistic simulation method for car-following models, which reasonably involves the stochastic nature of human driving behaviors and accurately replicates real-world traffic phenomena.
\end{enumerate}

This paper is organized as follows. Section~\ref{related_works} overviews the existing car-following models and simulation methods. Section~\ref{methods} introduces our proposed dynamic regression framework to address the limitations of existing models. Section~\ref{sec_exp} demonstrates the effectiveness of our model in calibration and simulation, followed by conclusions in Section~\ref{conclusion}.

\section{Related Works}\label{related_works}
Car-following models have been extensively used to understand and predict driver behavior in traffic, and model calibration is crucial. The traditional calibration methods, such as genetic algorithm-based calibration \citep{punzo2021calibration}, provide only point estimation for model parameters, lacking the ability to capture driving behavior uncertainty. On the contrary, probabilistic calibration and modeling approaches are commonly considered effective in addressing both epistemic uncertainties related to unmodeled details and aleatory uncertainty resulting from model prediction failures \citep{punzo2012can}.

A well-designed car-following model should capture both the inter-driver heterogeneity (diverse driving behaviors of different drivers \citep{ossen2006interdriver}) and intra-driver heterogeneity (the varying driving styles of the same driver \citep{taylor2015method}). Probabilistic calibration collects significant data from various drivers and conditions to fit the model parameters using statistical distributions instead of fixed values. For example, IDM parameters such as comfortable deceleration and maximum acceleration can be modeled as random variables with specific distributions \citep{treiber2017intelligent, zhang2024bayesian}, reflecting the range of driving styles. A typical probabilistic calibration method is maximum likelihood estimation (MLE) \citep{treiber2013traffic, zhou2023calibration}. Besides, using Fisher's information matrix \citep{spall2005monte} can also estimate the parameters' joint distribution via minimizing a certain objective function, since that it is the negative Hessian of the objective function at the estimated values and is the inverse of the covariance matrix for the joint distribution. Bayesian inference, as another probabilistic approach, combines prior knowledge and data to estimate the model parameter distribution, allowing for capturing inter-driver heterogeneity. By representing the behaviors probabilistically, a spectrum of plausible behaviors is obtained instead of a single deterministic response. Bayesian calibration approaches are discussed in detail in \cite{zhang2024bayesian}. To address intra-driver heterogeneity, stochastic car-following models are developed to account for a driver's behavior's dynamic, time-varying nature. These models introduce a random component that captures moment-to-moment behavior changes, such as using a stochastic process like Markov Chain to model a driver's reaction time or attention level. For instance, \cite{zhang2022generative} modeled each IDM parameter as a stochastic process and calibrated them in a time-varying manner.

In addition to traditional models, machine learning techniques have gained popularity in driving behavior modeling and prediction. These techniques leverage large datasets to capture complex, non-linear relationships and account for inter-driver and intra-driver heterogeneity. Data-driven models can be trained on a wealth of driving data to predict individual driver behavior and traffic flow, such as K-nearest neighbor algorithm (KNN) \citep{he2015simple}, Gaussian mixture model (GMM) \citep{zhang2024learning}, hidden Markov model (HMM) \citep{wang2018learning}, and long short-term memory (LSTM) networks \citep{wang2017capturing}. These models can complement traditional car-following models and enhance their predictive capabilities. Furthermore, the real-time adaptation of model parameters based on the current driving context and state of the driver and the vehicle can capture intra-driver heterogeneity. This involves continuous parameter updates using sensory information to reflect changes in driving conditions \citep{sun2022fast}.

Overall, probabilistic car-following models, enhanced by machine learning and real-time adaptation, offer a robust framework for capturing the observed diverse and complex driving behaviors. These models effectively handle the heterogeneity and uncertainty inherent in driving behavior by leveraging various methodologies. However, each model has its limitations. While data-driven models can capture complex patterns, they may lack interpretability and be data-hungry. These models are typically designed and trained to make predictions based on patterns they've learned from their training data. When they encounter data that falls outside of this distribution, their predictions can become unreliable or inaccurate, known as the ``out-of-distribution problem''. Recently, more advanced machine learning models (e.g., deep reinforcement learning \citep{hart2021formulation}) are trying to address these limitations. Traditional car-following models, though transparent, may not be flexible enough to encompass the full complexity of human driving behaviors. This work aims to leverage the prior knowledge of traditional car-following models while incorporating the flexibility needed to represent diverse human driving behaviors accurately. The efficacy of this approach is explicitly demonstrated through calibration and simulations of the IDM.

\section{Methodology}\label{methods}
\begin{figure}[t]
    \centering
    \includegraphics[width = 0.97\linewidth]{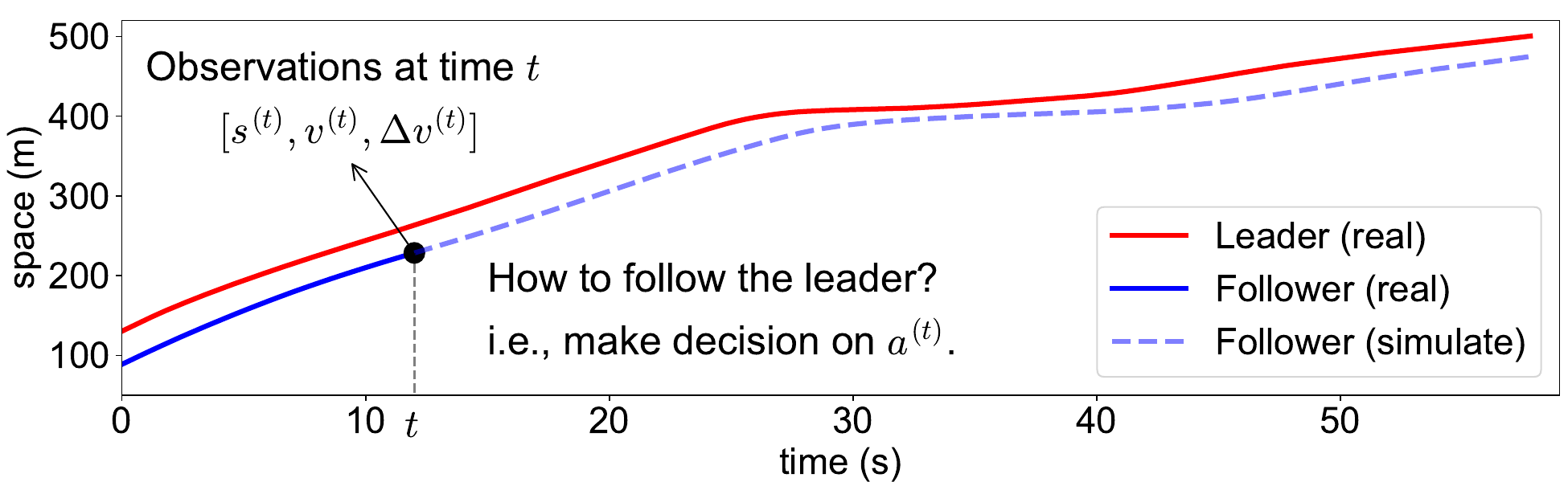}
    \caption{Illustration of the car-following problem in the time-spacing diagram.}
    \label{cf_illustration}
\end{figure}

\subsection{Preliminaries: IDM and Its Variants}
In this part, we assume that the internal states of each driver remain stationary, implying that their driving styles do not change within the region of interest (ROI). Thus we ignore the discussion of intra-driver heterogeneity. This assumption allows us to use a single model with time-invariant parameters to learn observed car-following behaviors. As a starting point, we introduce several models: the basic IDM with a deterministic formulation \citep{treiber2000congested}, the probabilistic IDM with action uncertainty \citep{treiber2017intelligent}, the Bayesian IDM with prior knowledge, and the memory-augmented IDM with a temporal structure \citep{zhang2024bayesian}.

\subsubsection{Intelligent Driver Model}
IDM \citep{treiber2000congested} is a continuous nonlinear function $f: \mathbb{R}^3 \mapsto \mathbb{R}$ which maps the gap, the speed, and the speed difference (approach rate)  to acceleration at a certain timestamp. Let $s$ represent the gap between the following vehicle and the leading vehicle, $v$ denote the following vehicle's speed, and $\Delta v=-ds/dt$ indicate the speed difference. The physical meanings of these notations are illustrated in \Cref{cf_illustration}, where $v_l$ denotes the speed of the leading vehicle. IDM computes vehicle acceleration using the following nonlinear function $f$:
\begin{align}\label{IDM_eq}
\begin{split}
    f(s,v,\Delta v)  \triangleq \alpha\,\left(1-\left({\frac{v}{v_{0}}}\right)^{4}-\left({\frac{s^{\ast}(v,\Delta v)}{s}}\right)^{2}\right),
\end{split}\\
\begin{split}\label{IDM_eq_gap}
  s^{\ast}(v,\Delta v) = s_{0}+ s_1\,\sqrt{\frac{v}{v_0}} + v\,T+{\frac{v\,\Delta v}{2\,{\sqrt{\alpha\,\beta}}}},
\end{split}
\end{align}
where $v_0,  s_{0}, T,  \alpha,$ and $\beta,$ are model parameters with specific physical meanings. The desired speed $v_{0}$ is the free-flow speed. The jam spacing $s_{0}$ denotes a minimum gap distance from the leading vehicle. The safe time headway $T$ represents the minimum interval between the following and leading vehicles. The acceleration $\alpha$ and the comfortable braking deceleration $\beta$ are the maximum vehicle acceleration and the desired deceleration to keep safe, respectively. The deceleration is controlled by the desired minimum gap $s^\ast$, and we set $s_1=0$ following \cite{treiber2000congested} to obtain a model with interpretable and easily measurable parameters.

\begin{figure}
    \centering
    {\centering
    \resizebox{0.65\linewidth}{!}{\input{Figs/decision_scheme}}}
    \caption{The decision-action simulation scheme.}
    \label{fig:decision_scheme}
\end{figure}
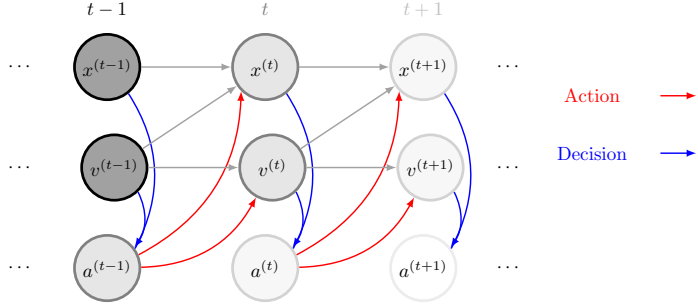

To make the notations concise and compact, we define a vector $\boldsymbol{\theta}=[v_0,s_0,T,\alpha,\beta] \in \mathbb{R}^5$ as the IDM parameters. For a certain driver $d$, we formulate the IDM acceleration term as $f(s_d^{(t)},v_d^{(t)},\Delta v_d^{(t)};\boldsymbol{\theta}_d)$. Compactly, we write the inputs at time $t$ as a vector $\boldsymbol{h}_d^{(t)}=[s_d^{(t)},v_d^{(t)},\Delta v_d^{(t)}]$. Then, we denote the IDM term as $\mathrm{IDM}(\boldsymbol{h}_d^{(t)};\boldsymbol{\theta}_d)$, further abbreviated as $\mathrm{IDM}_d^{(t)}$, where the subscript $d$ represents the index for each driver and the superscript $(t)$ indicates the timestamp. Given $\mathrm{IDM}_d^{(t)}$, we can update the vehicle speed $v^{(t+1)}$ and position $x^{(t+1)}$ following the ballistic integration scheme as in \cite{treiber2013microscopic} with a step of $\Delta t$:
\begin{subequations}\label{simulation_update}
    \begin{align}
        v^{(t+1)} &= v^{(t)} + a^{(t)} \Delta t,\label{update_v}\\
        x^{(t+1)} &= x^{(t)} + v^{(t)}\Delta t + \frac{1}{2}a^{(t)} \Delta t^2.\label{update_s}
    \end{align}
\end{subequations}
Specifically, three steps are involved in the discrete decision-making processes simulation shown in \Cref{simulation_update} and \Cref{fig:decision_scheme}. (i) Initialization: the available information at time $t$ includes $a^{(t-1)}$, $x^{(t)}$, and $v^{(t)}$; (ii) Decision-making: we estimate a possible action $\mathrm{IDM}^{(t)}$ based on IDM; (iii) Action execution and state updates: We take a specific action $a^{(t)}$ according to the decisions, resulting in the updated motion states $v^{(t+1)}$ (\Cref{update_v}) and $x^{(t+1)}$ (\Cref{update_s}). The calibration of IDM can be performed using different data, such as spacing, speed, and acceleration, as outlined in \cite{punzo2021calibration}. In this section, we focus on introducing the generative processes of acceleration data and provide detailed calibration methods on the speed or/and spacing in Section~\ref{dynamic_idm}.

\subsubsection{Probabilistic IDM with \emph{i.i.d.} Errors}
For the driver's aspect, imperfect and irregular driving behaviors result in erratic components of the driver's action \citep{treiber2013traffic, saifuzzaman2014incorporating}. Besides, some neglected aspects in the traditional car-following models may also contribute to the errors. For instance, more information from the leader (e.g., brake light signals), the ego vehicle (e.g., driver's action inertia), and the follower of the ego vehicle (e.g., nudging behaviors \citep{li2024modular}).
We can introduce some action noises with standard deviations $\sigma_{\eta}$ that are tolerated in the IDM framework to model such behaviors. Here, $\mathrm{IDM}_d$ represents the rational behavior model, while the random term accounts for the imperfect driving behaviors that IDM cannot capture. Considering the random term with the assumption of \textit{i.i.d.} noise, a probabilistic IDM \citep{bhattacharyya2020online, treiber2017intelligent} can be developed as
\begin{equation}\label{iid_a}
    \hat{a}_d^{(t)}|\boldsymbol{h}_d^{(t)},\boldsymbol{\theta}_d\, \stackon{$\sim$}{\tiny{\textit{i.i.d.}}}\, \mathcal{N}(\mathrm{IDM}_d^{(t)},\sigma_{\eta}^2),
\end{equation}
where $\mathrm{IDM}_d^{(t)}$ and $\sigma^2_{\eta}$ represent the mean and variance and $\hat{a}_d^{(t)}$ is the observed data of the true acceleration ${a}$.

The car-following model calibration is to estimate the model parameters $\boldsymbol{\theta}$ by seeking the best mapping from $\boldsymbol{h}_d^{(t)}$ to $a_d^{(t)}$, ${v}_d^{(t+1)}$, and $s_d^{(t+1)}$ \citep{punzo2021calibration}. Various approaches, such as utility-based optimization (e.g., genetic algorithm) and maximum-likelihood techniques, are usually adopted (see \cite{treiber2013microscopic}).

\subsubsection{Bayesian IDM with \emph{i.i.d.} Errors}
\begin{figure*}[t]
    \centering
    \subfigure{
        \centering  \includegraphics[width=.48\textwidth]{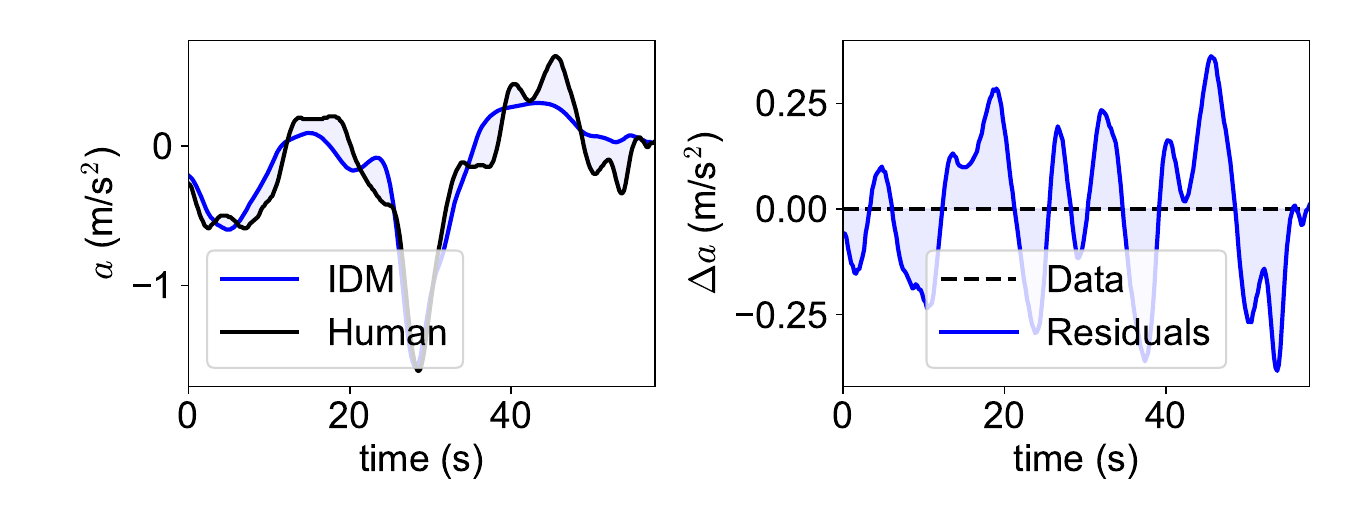}
    }
    \subfigure{
        \centering\includegraphics[width=.48\textwidth]{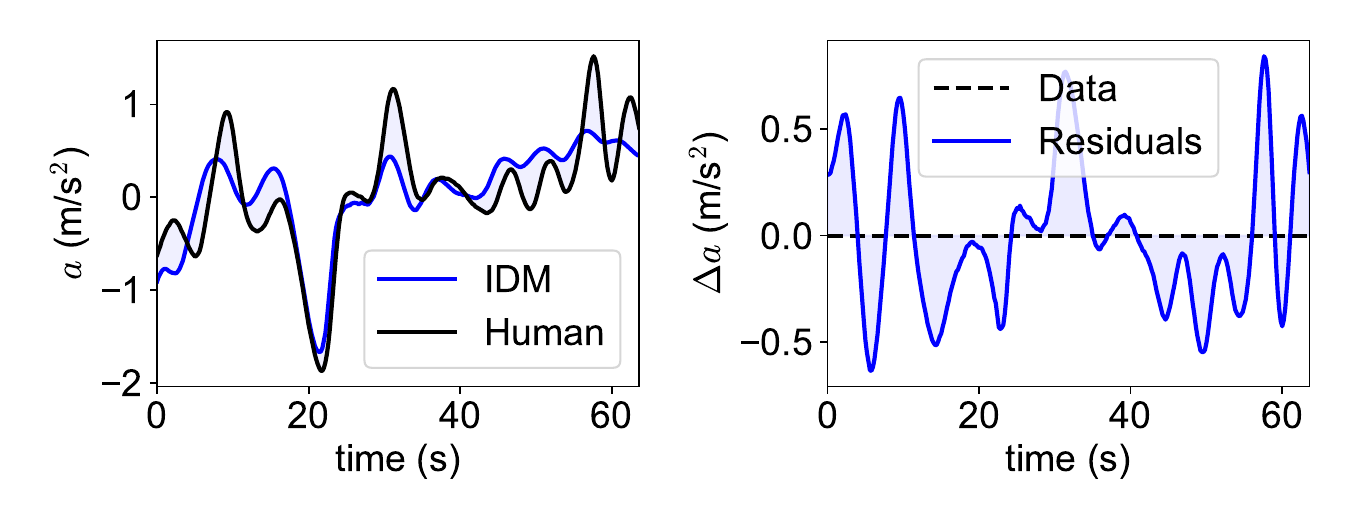}
    }\\
    \subfigure{
        \centering  \includegraphics[width=.48\textwidth]{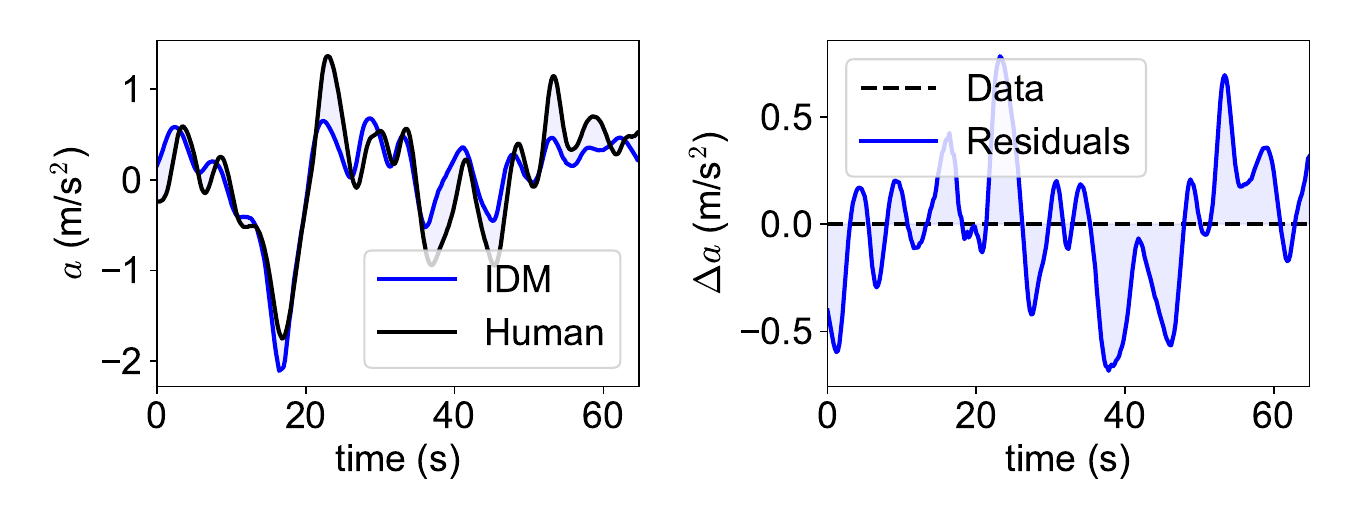}
    }
    \subfigure{
        \centering\includegraphics[width=.48\textwidth]{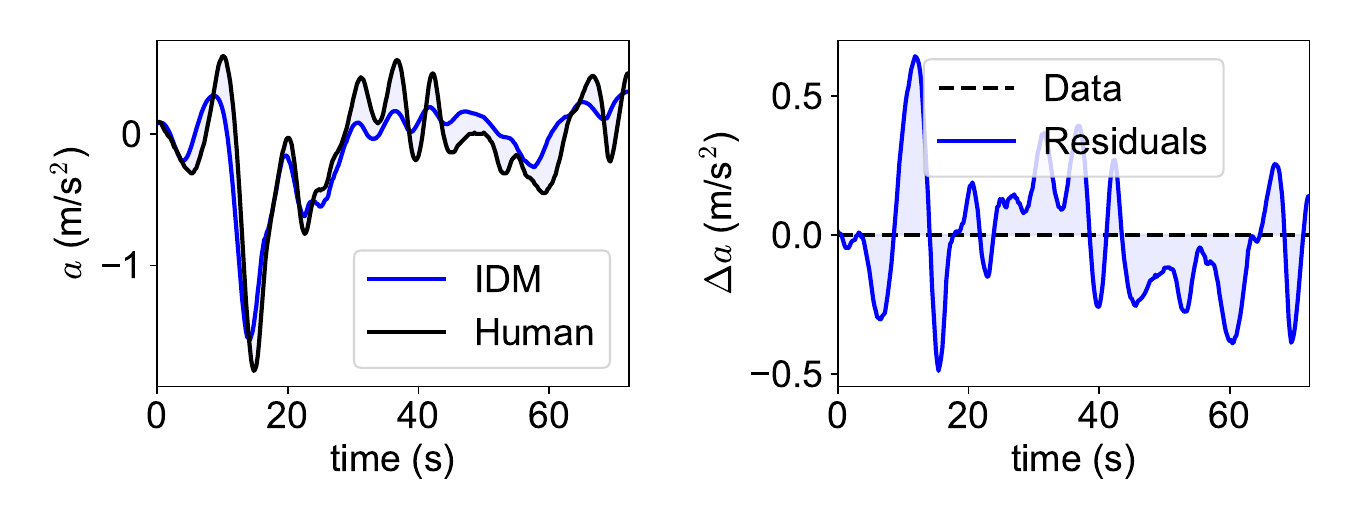}
    }
    \caption{Examples of the \textit{non-i.i.d.} residuals of the Bayesian IDM for four leader-follower pairs, where the left ones indicate the actual acceleration profiles v.s. the IDM term with identified parameters, the right ones are the residuals between the real data and the IDM term.}
    \label{fig:noniid}
\end{figure*}

We introduce a hierarchical Bayesian IDM, as proposed by \cite{zhang2024bayesian}, designed to capture general driving behaviors at the population level while also accounting for individual-level heterogeneity. This model is structured as follows: For any pair of time step and vehicle $(t,d)$ within the set $\{(t,d)\}_{t=t_0, d=1}^{t_0+T\Delta t,D}$, we have
\begin{subequations}
\begin{align}
    &\boldsymbol{\sigma}_0  \,\stackon{$\sim$}{\tiny{\textit{i.i.d.}}}\, \mathrm{Exp}(\lambda_0),\label{5a}\\
    &\boldsymbol{\Sigma}  \sim \mathrm{LKJCholeskyCov}(\eta, \boldsymbol{\sigma}_0), \label{5b}\\
    &\ln(\boldsymbol{\theta})  \sim \mathcal{N}(\ln(\boldsymbol{\theta}_{\text{rec}}), \boldsymbol{\Sigma}_0),\label{5c}\\
    &\sigma_{\eta} \sim \mathrm{Exp}(\lambda_{\eta}), \label{5e}\\
    & \mathbf{for~driver}\,d = 1,\ldots,D:\nonumber\\
    & \quad \quad \ln(\boldsymbol{\theta}_d)  \sim \mathcal{N}(\ln(\boldsymbol{\theta}), \boldsymbol{\Sigma}), \label{5d}\\
    & \quad \quad \mathbf{for~time}\, t=t_0,\ldots, t_0+(T_d-1)\Delta t:\nonumber\\
    &\quad \quad \quad \quad\hat{a}_d^{(t)}|\boldsymbol{h}_d^{(t)},\boldsymbol{\theta}_d\,  \stackon{$\sim$}{\tiny{\textit{i.i.d.}}}\, \mathcal{N}(\mathrm{IDM}_d^{(t)}, \sigma_{\eta}^2). \label{eq:bayesian_idm}
\end{align}
\end{subequations}
In this model, `LKJ' represents the LKJ distribution \citep{lewandowski2009generating}, and the hyperparameters $\lambda_0$, $\eta$, $\boldsymbol{\Sigma}_0$, and $\lambda_\eta$ are manually set, while other variables are inferred from the data. \Cref{5a} defines $\boldsymbol{\sigma}_0$ as a vector with components independently drawn from $\mathrm{Exp}(\lambda_0)$, forming the diagonal elements of a matrix. Then, given $\lambda_0$ and $\eta$, \Cref{5a} and \Cref{5b} establish the prior for the covariance matrix $\boldsymbol{\Sigma}\in \mathbb{R}^5$ of the IDM parameters, which modulates the model's hierarchical level between the pooled model and unpooled model. In a pooled model, all drivers are assumed to exhibit identical driving behaviors. Conversely, an unpooled model assumes each driver $d$ has a distinct, unrelated parameter set $\boldsymbol{\theta}_d$, as emphasized by \cite{zhang2024bayesian}. Then \Cref{5c} sets the prior for the population-level IDM parameters $\boldsymbol{\theta}$, and based on which \Cref{5d} describes the prior of the individual-level IDM parameters $\boldsymbol{\theta}_d$ for driver $d$. The hyperparameter $\lambda_\eta$ specifies the prior for the variance of the random noise in \Cref{5e}.
However, it is important to note that \Cref{eq:bayesian_idm} assumes \emph{i.i.d.} errors, which may not align with practical scenarios where residuals are often temporally correlated (\textit{non-i.i.d.}), as illustrated in \Cref{fig:noniid}.

\subsubsection{Memory-Augmented IDM with Gaussian Processes and \emph{i.i.d.} Errors}
Maintaining temporal consistency in actions in daily driving tasks is crucial for human drivers, commonly referred to as \emph{driving persistence} \citep{treiber2013traffic}. However, most stochastic models, including the Bayesian IDM, overlook the persistence of acceleration noise. Instead, they incorrectly model its time dependence as white noise, assuming \emph{i.i.d.} errors, as shown in \Cref{iid_a}, which can be reformulated as\begin{equation}\label{reformulated_iid_a}
    a_d^{(t)} = \mathrm{IDM}_{d}^{(t)} + \epsilon_d^{(t)},\ \epsilon_d^{(t)}\,\stackon{$\sim$}{\tiny{\emph{i.i.d.}}}\,\mathcal{N}(0,\sigma_{\eta}^2).
\end{equation}
Actually, these models and calibrated parameters are only valid when the errors adhere to the \emph{i.i.d.} assumption, and the temporal correlation in residuals leads to biased estimation of model parameters.

To address this limitation, \cite{zhang2024bayesian} proposed a Bayesian calibration method that involves Gaussian processes (GPs) to depict the ``memory effect'' \citep{treiber2003memory}, named as memory-augmented IDM (MA-IDM), which can capture the temporally correlated errors, $a_d^{(t)} = \mathrm{IDM}_{d}^{(t)} + a_{\mathrm{GP}\,d}^{(t)}$. Stacking the scalar $a_d^{(t)}$ along the time horizon $\forall t\in [t_0+\Delta t, t_0+T\Delta t]$ results in a vector $\boldsymbol{a}_{d}\in \mathbb{R}^{T}$. Such that we can derive a vector form $\boldsymbol{a}_{d} = \mathrm{\textbf{IDM}}_d + \boldsymbol{a}_{\mathrm{GP}\,d}$, with $\boldsymbol{a}_{\mathrm{GP}\,d}\sim \mathcal{N}(\boldsymbol{0}, \boldsymbol{K}_d)$, where $\boldsymbol{K}_d$ represents the kernel matrix constructed by the Squared-Exponential (SE) kernel with $\sigma_k$ and $\ell$. Then, based on and reuse \Cref{5a,5b,5c,5d}, the MA-IDM is structured as
\begin{subequations}
\begin{align}
    &\sigma_{k}  \sim \mathrm{Exp}(\lambda_{k}),\\
    &\ln(\ell)  \sim \mathcal{N}(\ln(\mu_{\ell}),\sigma_{\ell_0}^2), \\
    & \mathbf{for~driver}\,d = 1,\ldots,D:\nonumber\\
    &\quad \quad \ln(\sigma_{k,d})  \sim \mathcal{N}(\ln(\sigma_{k}),\sigma_{\sigma}^2) \in \mathbb{R}, \\
    &\quad \quad \ln(\ell_{d})  \sim \mathcal{N}(\ln(\ell),\sigma_{\ell}^2) \in \mathbb{R}, \\
    &\quad \quad {\boldsymbol{a}}_d|\boldsymbol{h}_d,\boldsymbol{\theta}_d\, \stackon{$\sim$}{\tiny{\textit{i.i.d.}}}\, \mathcal{N}\left(\mathrm{\textbf{IDM}}_d,\boldsymbol{K}_d\right).
\end{align}
\end{subequations}

\subsection{Calibration of the Dynamic IDM with Autoregressive Processes and \emph{i.i.d.} Errors}\label{dynamic_idm}

The kernel function selection restricts the flexibility of modeling the temporally correlated errors with GPs. To address this limitation, we propose a novel unbiased hierarchical Bayesian model called dynamic IDM. This model addresses this limitation by incorporating AR processes to represent the time-dependent stochastic error term with a dynamic regression framework. By absorbing AR processes in the calibration framework, we adopt a simple yet effective approach of adding up actions along the history steps. This approach has been tested and proven effective in modeling car-following behaviors \citep{ma2005dynamic}.

Here we distinguish the concepts of ``states'' and ``observations'', where states (i.e., the real acceleration) represent the underlying dynamics of the vehicle, and observations are the measured values (e.g., speed and spacing). In \Cref{model_AR}, we first build the equation of states, which describes how the system evolves along the time horizon. Then in the rest of this subsection, we develop the observation equation, which describes how the underlying state is transformed (with noise added) into something that we directly measure.

\subsubsection{Modeling the temporal correlations with AR processes}\label{model_AR}
In our model, we conceptualize data error as comprising two distinct parts: process error and observation/measurement noise. The process error accounts for the inherent variability and uncertainties in the car-following process itself, which are not directly observable but have a significant impact on the model's behavior. The observation or measurement noise, on the other hand, is typically assumed to be independent and identically distributed (\textit{i.i.d.}). This assumption is based on the nature of measurement errors, which are often random and uncorrelated.

Specifically, we assume the process error (i.e., error of the states) follows a $p$-order AR process denoted by $\mathrm{AR}(p)$:
\begin{align}\label{a_AR}
    a_d^{(t)} &= \mathrm{IDM}_{d}^{(t)} + \epsilon^{(t)}_d,\\
    \epsilon^{(t)}_d &= \rho_{d,1}\epsilon_{d}^{(t-1)} + \rho_{d,2}\epsilon_{d}^{(t-2)}+\dots + \rho_{d,p}\epsilon_{d}^{(t-p)} + \eta_{d}^{(t)},
\end{align}
where $\eta^{(t)} \,\stackon{$\sim$}{\tiny{\textit{i.i.d.}}} \mathcal{N}(0,\sigma_\eta^2)$ represents a white noise series. In the following discussions, we refer to $\mathrm{IDM}_{d}^{(t)}$ as the \textit{mean component} and $\epsilon^{(t)}_d$ as the \textit{stochastic component}.

We estimate the model (\Cref{a_AR}) by constructing the likelihood on a white noise process $\eta_{d}^{(t)}$. Given the model with estimated parameters, the probabilistic prediction can be achieved by first sampling $\eta_{d}^{(t)}$ and then sequentially feeding it into
\begin{align}
    a_d^{(t)} =& \, \mathrm{IDM}_d^{(t)}+ \rho_{d,1}(a_d^{(t-1)} - \mathrm{IDM}_d^{(t-1)}) + \rho_{d,2}(a_d^{(t-2)} - \mathrm{IDM}_d^{(t-2)})\nonumber+ \cdots\\ &\,+ \rho_{d,p}(a_d^{(t-p)} - \mathrm{IDM}_d^{(t-p)}) +\eta_{d}^{(t)}.
\end{align}
The above equation explicitly demonstrates the advantages of our method compared with traditional models --- \textbf{it involves rich information from several historical steps to make decisions for the current step instead of using only one historical step}.

Here, we introduce the form of the hierarchical dynamic IDM as
\begin{subequations}
\begin{align}
    &\sigma_{\eta}  \sim \mathrm{Exp}(\lambda_{\eta}), \label{12e}\\
    &\boldsymbol{\rho} \sim \mathcal{N}(\boldsymbol{0}, \sigma_{\rho_0}^2\boldsymbol{I}), \label{12f}\\
    & \mathbf{for~driver}\,d = 1,\ldots,D:\nonumber\\
    &\quad \quad \boldsymbol{\rho}_d \sim \mathcal{N}(\boldsymbol{\rho}, \sigma_\rho^2\boldsymbol{I}),\\
    & \quad \quad \mathbf{for~time}\, t=t_0,\ldots, t_0+(T_d-1)\Delta t:\nonumber\\
    &\quad \quad \quad \quad {a}_d^{(t)}|\boldsymbol{h}_d^{(t)},\boldsymbol{\theta}_d\,  \stackon{$\sim$}{\tiny{\textit{i.i.d.}}}\, \mathcal{N}\left(\mathrm{IDM}_d^{(t)}+\sum_{k=1}^{p}\rho_{d,k} \left(a_d^{(t-k)}-\mathrm{IDM}_d^{(t-k)} \right) , \sigma_{\eta}^2 \right),\label{12g}
\end{align}
\end{subequations}
where $\boldsymbol{\rho}=[\rho_1,\rho_2,\dots,\rho_p]$ and $\boldsymbol{\rho}_d=[\rho_{d,1},\rho_{d,2},\dots,\rho_{d,p}]$ are estimated during model training. The priors of the AR coefficients are normally distributed with zero means, as shown in \Cref{12f}. We reuse the same distribution in \Cref{5a,5b,5c,5d,5e} for the other variables. The likelihood can then be derived as a normal distribution shown in \Cref{12g}.

It is worth noting that the Bayesian IDM is a special case of the dynamic IDM. If we set the AR order as zero (i.e., $p=0$), this model will be exactly equivalent to the Bayesian IDM. Besides, when the AR order $p=1$, the dynamic IDM can be thought of as a discrete-time analogue to the model proposed by \cite{treiber2006delays}, where the correlated acceleration noise obeys an Ornstein-Uhlenbeck (OU) process, which is the continuous version of AR(1). Furthermore, in MA-IDM, a GP with a Matérn 1/2 kernel can also be seen as the continuous-time limit of an AR(1) process. All of these models share a key similarity in the type of correlation they model, i.e., exponential decay of correlation.

\subsubsection{Calibration on speed data with observation noise}\label{calibrate_AR1}
Our proposed method can be used for the calibration based on the acceleration, speed, and spacing data. Here, we provide a brief introduction to the calibration method based on speed data. More details can be found in the appendix.

According to \Cref{update_v}, the likelihood of a noisy speed observation is written as
\begin{equation}
    \hat{v}_{d}^{(t+1)}\sim \mathcal{N}\left(\bar{v}_{d}^{(t+1)}, (\sigma_\eta\Delta t)^2+\sigma_v^2\right),
\end{equation}
where the mean can be written as
\begin{equation}
    \bar{v}_{d}^{(t+1)}=v_{d}^{(t)}+ \mathrm{IDM}_{d}^{(t)}\Delta t+ \sum_{k=1}^p \rho_{d,k} \left(v_{d}^{(t-(k-1))} - v_{d}^{(t-k)}\right) - \sum_{k=1}^p \rho_{d,k} \mathrm{IDM}_{d}^{(t-k)}\Delta t,\nonumber
\end{equation}
the observation $\hat{v}_{d}^{(t+1)}$ is the measured data of the true speed ${v}_{d}^{(t+1)}$, and $\sigma_v^2$ is the variance of the observation noise.

\subsubsection{Calibration on position/spacing data with observation noise}\label{calibrate_AR2}
According to \Cref{update_s}, the likelihood of a noisy positional observation is written as
\begin{equation}
    \hat{x}_{d}^{(t+1)} \sim \mathcal{N}\left(\bar{x}_{d}^{(t+1)}, \Bigl(\frac{1}{2}\sigma_\eta\Delta t^2\Bigr)^2+\sigma_x^2\right),
\end{equation}
where the mean can be written as
\begin{equation}
    \bar{x}_{d}^{(t+1)}=x_{d}^{(t)} + v_{d}^{(t)}\Delta t + \frac{1}{2}\mathrm{IDM}_{d}^{(t)}\Delta t^2 + \frac{1}{2}\sum_{k=1}^p \rho_{d,k} \left(v_{d}^{(t-(k-1))} - v_{d}^{(t-k)}\right)\Delta t  - \frac{1}{2}\sum_{k=1}^p \rho_{d,k} \mathrm{IDM}_{d}^{(t-k)}\Delta t^2,\nonumber
\end{equation}
the observation $\hat{x}_{d}^{(t+1)}$ is the measured data of the true position ${x}_{d}^{(t+1)}$, and $\sigma_x^2$ is the variance of the observation noise of position data. Note that this is a position-based form, but one can easily adapt it into a gap-based form.

\subsubsection{Calibration on both speed and position data with joint likelihood}\label{calibrate_AR3}
By jointly considering the information and observation noise from the speed and position data, we can derive the likelihood using a bi-variate normal distribution, written as

\begin{equation}
    \left[\begin{array}{c}
         \hat{x}_{d}^{(t+1)}  \\
         \hat{v}_{d}^{(t+1)}
    \end{array}\right] \sim \mathcal{N}\left(\left[\begin{array}{c}
         \bar{x}_{d}^{(t+1)}  \\
         \bar{v}_{d}^{(t+1)}
    \end{array}\right], \underbrace{\left[\begin{array}{cc}
         \Bigl(\frac{1}{2}\sigma_\eta\Delta t^2\Bigr)^2 & \frac{1}{2}\sigma_\eta^2\Delta t^3 \\
         \frac{1}{2}\sigma_\eta^2\Delta t^3 & (\sigma_\eta\Delta t)^2
    \end{array}\right]}_{\text{process error}} + \underbrace{\left[\begin{array}{cc}
         \sigma_x^2 & 0 \\
         0 & \sigma_v^2
    \end{array}\right]}_{\text{observation noise}} \right).
\end{equation}
The variance values  ($\sigma_x$ and $\sigma_v$) in the observation noise determine the reliability of the position and speed data. When $\sigma_x$ is with a large value, this form tends to be equal to pure calibration on the speed data. Similarly, when $\sigma_v$ is large, it equals calibration only using the position data. To let our model automatically identify different noise levels, it is necessary to add two more priors on $\sigma_v$ and $\sigma_x$ as
\begin{align}
    \sigma_{v} & \sim \mathrm{Exp}(\lambda_{v}) \in \mathbb{R}^+,\\
    \sigma_{x} & \sim \mathrm{Exp}(\lambda_{x}) \in \mathbb{R}^+.
\end{align}
Note that here we assume the measurement noises as \emph{i.i.d.}, but generally they are also temporally correlated. This assumption has no serious consequences since the correlated measurement noises can be absorbed into the latent process error term.

\section{Experiments and Simulations}\label{sec_exp}
Calibration of the car-following behaviors essentially is a regression task from the perspective of a machine learning practitioner. In this section, we evaluate the calibration results from two aspects, the regression results (e.g., \Cref{fig:noniid}) and the simulation performance. We begin by evaluating the regression results and analyzing the identified parameters of IDM and the learned AR processes. Next, we demonstrate the simulation capability of our proposed method through short-term and long-term simulations. The short-term simulations quantitatively assess the replication of human behaviors, while the long-term simulations validate the ability of the identified parameters to capture critical car-following dynamics.

\subsection{Experimental Settings}
\subsubsection{Dataset}
The car-following model performance can be affected by noise in empirical data \citep{montanino2015trajectory}. To mitigate the impact of data quality and avoid the need for excessive data filtering, selecting an appropriate dataset is crucial. Different datasets can be utilized to verify various aspects of model capability. For instance, studying general car-following behaviors based on deep learning models requires informative data with sufficiently long trajectories. Exploring both population-level and individual-level driving behaviors based on hierarchical models necessitates multi-user driving data. Similarly, analyzing the driving style shifts in a single driver based on behavior models relies on the daily trajectories of that specific driver.

To access hierarchical car-following models for multiple drivers, we evaluate our model using the HighD dataset \citep{highDdataset}, which offers high-resolution trajectory data collected with drones. The HighD dataset provides several advantages over the NGSIM dataset \citep{punzo2011assessment} with advanced computer vision techniques and more reliable data-capture methods. It consists of $60$ video recordings from several German highway sections, each spanning a length of $420$ m. The original dataset is downsampled to a smaller set with a sampling frequency of $5$ Hz, achieved by uniformly selecting every $5$-th sample. The recorded trajectories, speeds, and accelerations of two types of vehicles (car and truck) are measured and estimated. We follow the same data processing procedures as in \cite{zhang2021spatiotemporal} to transform the data into a new coordinate system. We prefer trajectories with car-following duration longer than a certain threshold $t_0 = 50$ s for robust estimation of IDM parameters \citep{punzo2014we}. Then, we randomly select $20$ leader-follower pairs from these data for each type of vehicle, respectively.

\begin{figure}[t]
    \centering
    \includegraphics[width = \linewidth]{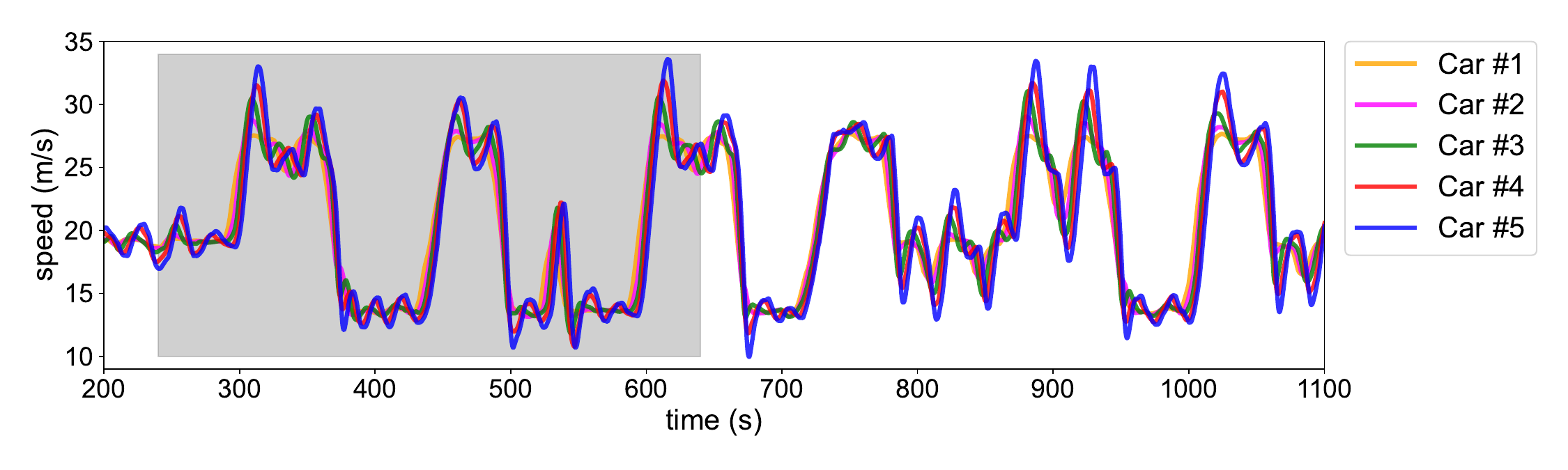}
    \caption{The illustration of $5$ vehicles' speed profiles in the OpenACC dataset, where the shaded area ($240\sim640$ s) is selected for the platoon simulation.}
    \label{openacc_data}
\end{figure}

Since the HighD dataset only provides short trajectories with duration up to about $ 60$ s, we also evaluate our method on OpenACC\footnote{\url{https://data.jrc.ec.europa.eu/dataset/9702c950-c80f-4d2f-982f-44d06ea0009f}.} trajectories with a duration of about $1000$ s. OpenACC dataset provides car-following records of a platoon, where a human driver operates the first vehicle, and four following vehicles in the platoon are autonomously controlled by an adaptive cruise control (ACC) module, as shown in \Cref{openacc_data}. The data in OpenACC is also downsampled to $5$ Hz.

\begin{table*}[t]
\tiny
\centering
\caption{Posterior Mean of Model Parameters.}
\label{posterior_expectations_table}
\begin{tabular}{cccl}
\toprule
Models & $\boldsymbol{\theta}=[v_0,s_0,T,\alpha,\beta]$  & $\sigma_\eta$ & $\rho$      \\ \midrule
MA-IDM & $[16.919,3.538,1.183,0.553,2.147]$  & $\slash$    & ($\sigma_k=0.202$, $\ell=1.44$ s) \\
Bayesian IDM ($p=0$) & $[21.090, 3.724, 0.946, 0.518, 1.542]$ & $0.240$    & $\slash$ \\
Dynamic IDM ($p=1$) & $[29.738,3.220,1.186,0.769,4.130]$ & $0.019$  & $[0.989]$ \\
Dynamic IDM ($p=2$) & $[27.592,3.367,1.191,0.741,3.483]$ & $0.019$  & $[1.234,-0.247]$ \\
Dynamic IDM ($p=3$) & $[25.004,2.974,1.206,0.811,2.442]$ & $0.017$  & $[1.123,0.425,-0.572]$ \\
Dynamic IDM ($p=4$) & $[26.181,2.850,1.222,0.811,3.145]$ & $0.016$  & $[0.901,0.590,-0.149,-0.377]$ \\
Dynamic IDM ($p=5$) & $[27.099,2.843,1.235,0.813,3.422]$ & $0.016$  & $[0.874,0.580,-0.105,-0.315,-0.071]$ \\
Dynamic IDM ($p=6$) & $[28.089,2.702,1.259,0.826,3.325]$ & $0.015$  & $[0.902,0.632,-0.100,-0.427,-0.217,0.181]$ \\
Dynamic IDM ($p=7$) & $[28.574,2.594,1.276,0.817,3.439]$ & $0.014$  & $[0.866,0.690,-0.001,-0.413,-0.378,-0.032,0.248]$ \\
Dynamic IDM ($p=8$) & $[28.446,2.573,1.264,0.796,3.805]$ & $0.014$  & $[0.816,0.700,0.075,-0.331,-0.381,-0.172,0.080,0.200]$ \\
\bottomrule
\multicolumn{4}{l}{* Recommendation values \citep{treiber2000congested}: $\boldsymbol{\theta}_{\text{rec}}=[33.3, 2.0, 1.6, 1.5, 1.67]$.}
\end{tabular}
\end{table*}

\subsubsection{Model Training}
We develop our model using PyMC 4.0 \citep{salvatier2016probabilistic}. Codes for all experimental results reported in this paper are released in \url{https://github.com/Chengyuan-Zhang/IDM_Bayesian_Calibration}. Specifically, we adopt the No-U-Turn Sampler, a kind of Hamiltonian Monte Carlo method \citep{hoffman2014no}, and set the burn-in steps as $3000$ to ensure the detailed balance condition holds for the Markov chains. We recommend and utilize the values $\boldsymbol{\theta}_{\text{rec}}=[33.3, 2.0, 1.6, 1.5, 1.67]$ in \cite{treiber2000congested} for the IDM priors. In addition, we set $\lambda_0=100$ for the exponential distribution, $\Sigma_0=\mathrm{diag}([0.1, 0.1, 0.1, 0.1, 0.1])$, and $\eta=2$ for the LKJCholeskyCov distribution.

It's important to recognize that identifiability issues can emerge when distinguishing the individual contributions of different model components becomes challenging, particularly due to their overlapping effects on the output. This is evident in our case, where both the IDM and AR processes aim to explain similar aspects of driving behavior, thus complicating the task of isolating their distinct impacts. To mitigate this issue, we have implemented a strategy of setting a strong, yet narrowly defined prior on the AR term (i.e., $\lambda_\eta=2\times10^6$) and the error term (i.e., $\lambda_{v}=10^6$ and $\lambda_{x}=10^7$). This approach is designed to ensure that the IDM predominantly explains the information in the data, thereby minimizing the potential overlap and enhancing the clarity of each component's role in the model.

\begin{figure*}[t]
    \centering
    \subfigure[The Empirical covariance matrix.]{
        \centering
        \includegraphics[width=0.313\textwidth]{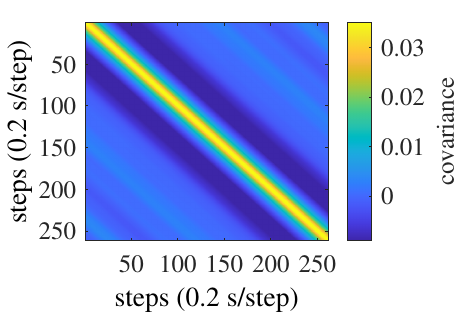}
    }
    \subfigure[The empirical covariance function.]{
        \centering
        \includegraphics[width=0.5\textwidth]{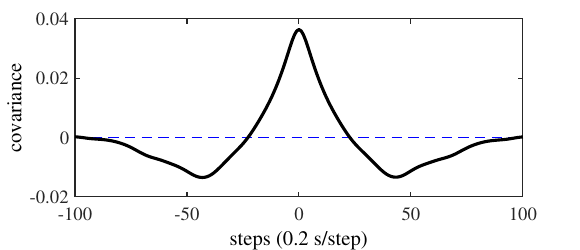}
    }
    \subfigure[The SE kernel matrix.]{
        \centering
        \includegraphics[width=0.313\textwidth]{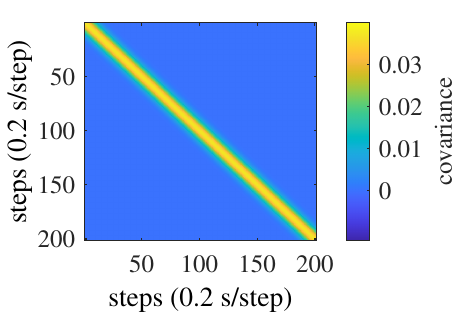}
    }
    \subfigure[The SE kernel function.]{
        \centering
        \includegraphics[width=0.5\textwidth]{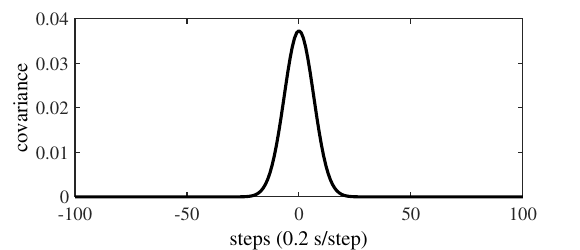}
    }
    \subfigure[The AR(5) covariance matrix.]{
        \centering
        \includegraphics[width=0.313\textwidth]{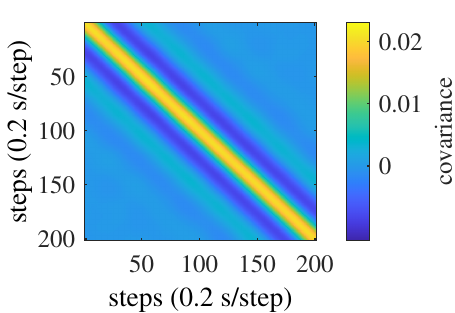}
    }
    \subfigure[The AR(5) covariance functions.]{
        \centering
        \includegraphics[width=0.5\textwidth]{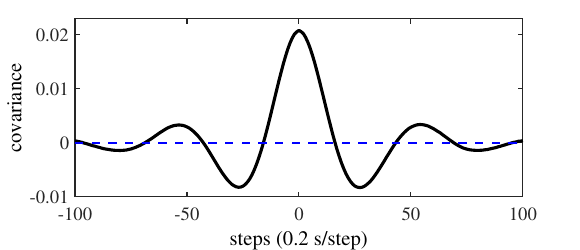}
    }
    \caption{The comparison among the empirical covariance, SE kernel, and AR processes.}
    \label{fig:RBFvsAR_cov}
\end{figure*}

\begin{figure*}[t]
    \centering
    \includegraphics[width=.95\linewidth]{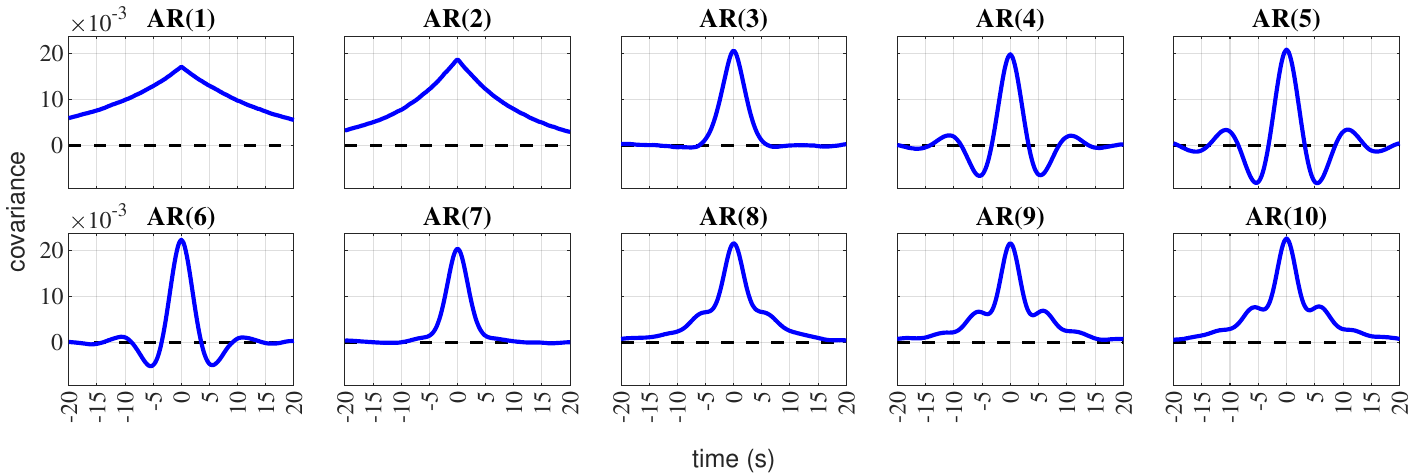}
    \caption{The AR covariance functions with different order $p$.}
    \label{fig:10cov_funcs}
\end{figure*}

\subsection{Calibration Results and Analysis}
We compare the parameters of the Bayesian IDM and the MA-IDM with the dynamic IDM in \Cref{posterior_expectations_table}. The parameter $\sigma_\eta$ represents the noise level of the \emph{i.i.d.} errors, which decreases as the AR order increases. This indicates that the residual of the Bayesian IDM (when $p=0$) retains valuable information, as reflected by a higher value of $\sigma_\eta$. With higher AR order, more information is captured by the AR processes, reducing the contribution of the noise term. It is worth noting that the model with AR ($p=0$) is equivalent to the Bayesian IDM.

To compare the roles of GP and AR processes, we examine the SE kernel matrix alongside the empirical covariance matrix estimated from real data and the covariance matrix generated from AR(5), see \Cref{fig:RBFvsAR_cov}. The horizontal axes are time steps, where each time step represents $0.2$ s. The SE kernel merely captures short-term positive correlations but fails to capture the long-term negative correlations observed in the residuals extracted from real driving data. To address this limitation, the dynamic IDM involves AR processes to capture both \textit{short-term positive} and \textit{long-term negative} correlations more realistically. This finding suggests that \textbf{human drivers' immediate stochastic component can be significantly impacted by their short-term  $\epsilon_d^{(t)}$ values (up to $5$ s)} \citep{zhang2024bayesian}, potentially due to driving persistence \citep{treiber2013traffic}. Moreover, we uncover a notable observation: \textbf{the stochastic components are also negatively correlated with the previous $\epsilon_d^{(t)}$ values in a long-term period ($5\sim10$ s)}. For example, if the acceleration value of the human driver's current action is greater than the expected behavior, i.e., the mean component in \Cref{a_AR}, then in the following $5\sim10$ s, the driver is very likely to take an action whose value is lower than the expected behaviors. We then compare the covariance functions generated by AR processes with different orders $p=1,\dots, 10$, as shown in \Cref{fig:10cov_funcs}. Generally, increasing the AR order $p$ enables better capture of the underlying structure in a time series. However, this comes with trade-offs. Too few lags constrain the power of AR processes, resulting in underfitting (e.g., AR(1)-AR(3)). Conversely, too many lags can lead to overfitting and impose a heavier computational burden (e.g., AR(7)-AR(10)). Therefore, the optimal AR order (e.g., AR(4)-AR(6)) should be determined and verified experimentally.

\subsection{Simulations}
Simulation of car-following models ensuring practical applicability and robustness under diverse conditions. One of the highlights of the dynamic IDM is its simulation scheme, leveraging the explicitly defined generative processes of the error term presented in \Cref{a_AR}. This feature allows for the simulation and updating of future motions based on these generative processes. In what follows, we will introduce the parameter generation method for IDM and develop a stochastic simulator, which has been successfully tested in multiple cases.

\subsubsection{Probabilistic Simulations}
\cite{zhang2024bayesian} emphasized a key advantage of the Bayesian method over genetic algorithm-based approaches. The Bayesian method provides joint distributions of parameters instead of only a single value. Consequently, instead of relying solely on the mean values from \Cref{posterior_expectations_table}, we can draw a large number of samples from the posterior joint distributions to generate anthropomorphic IDM parameters effectively.

\begin{figure*}[t]
    \centering
    \subfigure[Truck driver $\# 211$.]{
        \centering
        \includegraphics[width=0.48\textwidth]{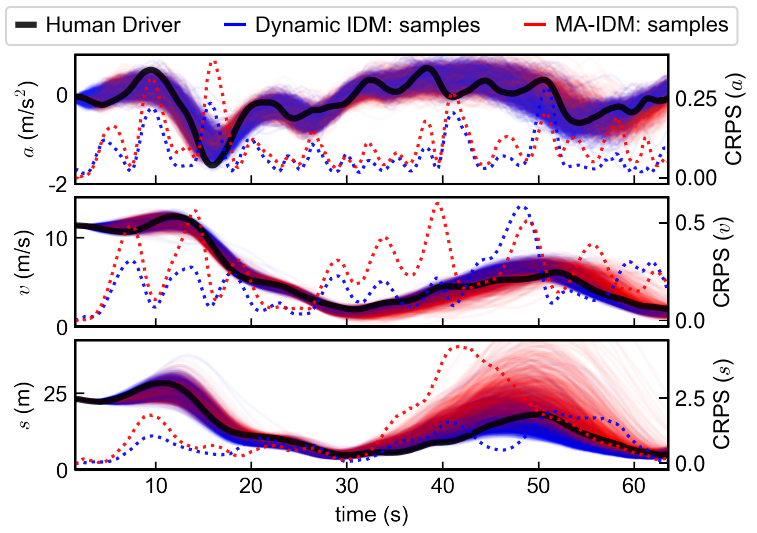}
    }
    \subfigure[Car driver $\# 273$.]{
        \centering
        \includegraphics[width=0.48\textwidth]{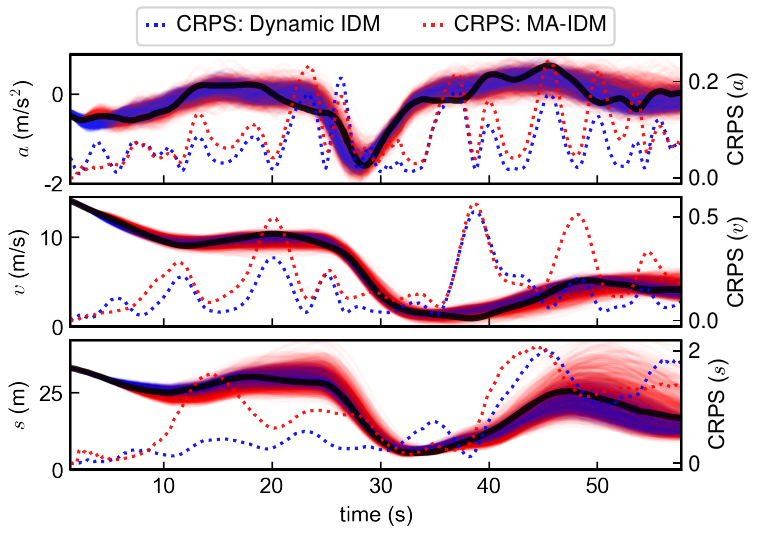}
    }
    \caption{The probabilistic simulation results of a representative truck driver. The black lines indicate the ground truth of human driving data; The red and blue lines are the predicted motion states with the parameter samples drawn from posteriors, while the dotted lines are the CRPS (see \Cref{crps}) at the corresponding time step.}
    \label{fig:simulation_for_two_cars}
\end{figure*}

\begin{figure}[t]
    \centering
    \includegraphics[width=.9\linewidth]{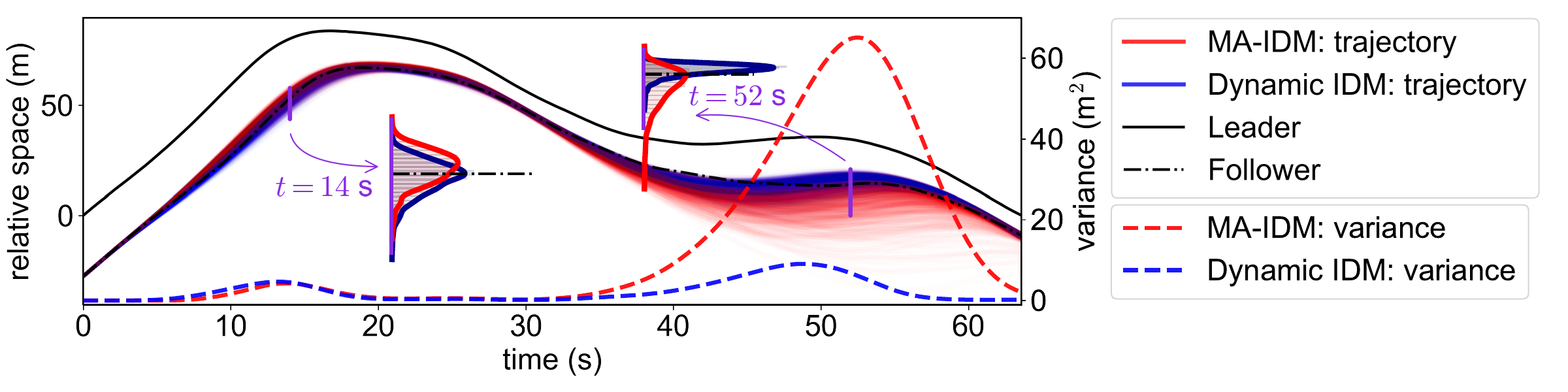}
    \caption{The time-space diagram of the $\# 211$ truck follower's posterior trajectories captured by another `observing' vehicle with a constant mean speed.}
    \label{fig:ts_diag}
\end{figure}

We select one driver as an illustrative simulation example to show the calibration results. We generate $N=1000$ sets of IDM parameters from the posteriors of the calibrated model. Then, given the leading vehicle's full trajectories and only the initial states of the follower, we simulate the follower's full trajectories. The follower's behavior is controlled by the calibrated IDM, with the addition of a random noise sampled from the generative processes. The simulation step is set as $0.2$ s. According to \Cref{a_AR}, the simulation process involves three steps: (1) generating the mean model by sampling a set of IDM parameters from $\theta_d$, (2) computing the serial correlation term according to the historical information, and (3) sampling white noise randomly.

\begin{table*}[t]
    \footnotesize
    \centering
    \caption{Evaluations of the short-term (5 s) simulations with different models. All values are amplified by ten times to keep an efficient form.}
    \begin{tabular}{c|c|c|c|c|c|c}
    \toprule
        = real values $\times 10$ &  $\mathrm{RMSE}(a)$  & $\mathrm{RMSE}(v)$ & $\mathrm{RMSE}(s)$ & ${\mathrm{CRPS}}(a)$ & ${\mathrm{CRPS}}(v)$ & ${\mathrm{CRPS}}(s)$ \\
    \midrule
        MA-IDM & $2.03 \pm 0.48$ & $3.00 \pm 0.59$ & $5.15 \pm 0.86$ & $1.11 \pm 0.32$ & $1.62 \pm 0.39$ & $3.14 \pm 0.58$ \\
        Bayesian IDM ($p=0$) & $3.19 \pm 0.62$ & $2.90 \pm 0.83$ & $6.00 \pm 1.83$ & $1.25 \pm 0.33$ & $1.92 \pm 0.62$ & $3.91 \pm 1.29$ \\
        Dynamic IDM ($p=1$) & $1.78 \pm 0.54$ & $2.83 \pm 0.87$ & $4.94 \pm 1.46$ & $1.26 \pm 0.44$ & $1.95 \pm 0.67$ & $3.07 \pm 0.98$ \\
        Dynamic IDM ($p=2$) & $1.74 \pm 0.44$ & $2.68 \pm 0.66$ & $4.78 \pm 1.14$ & $1.18 \pm 0.36$ & $1.77 \pm 0.48$ & $2.88 \pm 0.76$ \\
        Dynamic IDM ($p=3$) & $1.77 \pm 0.46$ & $2.77 \pm 0.79$ & $4.68 \pm 1.26$ & $1.10 \pm 0.35$ & $1.66 \pm 0.55$ & $2.51 \pm 0.79$ \\
        Dynamic IDM ($p=4$) & $1.76 \pm 0.55$ & $2.71 \pm 0.85$ & $4.43 \pm 1.29$ & $1.08 \pm 0.42$ & $1.64 \pm 0.61$ & $2.40 \pm 0.83$ \\
        Dynamic IDM ($p=5$) & $\mathbf{1.66 \pm 0.38}$ & ${2.65 \pm 0.66}$ & ${4.29 \pm 1.01}$ & $\mathbf{0.95 \pm 0.28}$ & $\mathbf{1.49 \pm 0.45}$ & $\mathbf{2.17 \pm 0.63}$ \\
        Dynamic IDM ($p=6$) & $1.76 \pm 0.51$ & $2.72 \pm 0.81$ & $4.41 \pm 1.22$ & $1.07 \pm 0.39$ & $1.60 \pm 0.56$ & $2.32 \pm 0.76$ \\
        Dynamic IDM ($p=7$) & $1.68 \pm 0.39$ & $2.65 \pm 0.68$ & $4.28 \pm 1.09$ & $1.00 \pm 0.29$ & $1.56 \pm 0.46$ & $2.24 \pm 0.69$ \\
        Dynamic IDM ($p=8$) & $1.68 \pm 0.46$ & $2.65 \pm 0.74$ & $4.27 \pm 1.11$ & $1.01 \pm 0.35$ & $1.55 \pm 0.51$ & $2.25 \pm 0.71$ \\
        Dynamic IDM ($p=9$) & $1.68 \pm 0.47$ & $\mathbf{2.63 \pm 0.77}$ & $\mathbf{4.23 \pm 1.17}$ & $1.01 \pm 0.35$ & $1.53 \pm 0.54$ & $2.23 \pm 0.75$ \\
        Dynamic IDM ($p=10$) & $1.72 \pm 0.46$ & $2.68 \pm 0.76$ & $4.27 \pm 1.07$ & $1.04 \pm 0.34$ & $1.56 \pm 0.52$ & $2.23 \pm 0.65$ \\
    \bottomrule
    \end{tabular}
    \label{tab:errors}
\end{table*}

\begin{figure*}[t]
    \centering
    \includegraphics[width=.85\textwidth]{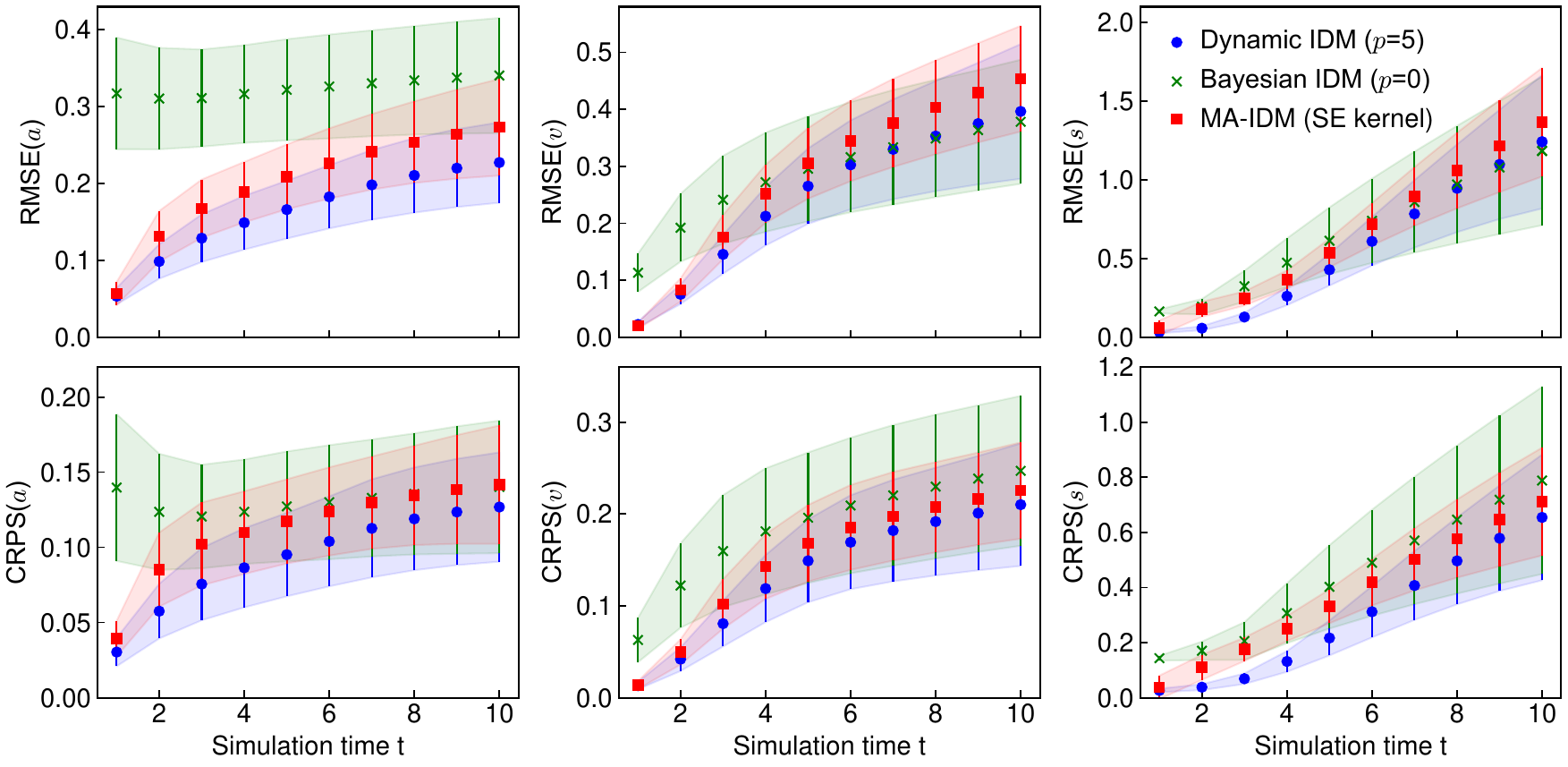}
    \caption{The performance comparison of 1 to 10 seconds simulations.}
    \label{fig:1to10_sim}
\end{figure*}

We compare our model with the baseline MA-IDM \citep{zhang2024bayesian}. Simulated trajectories are generated using parameter samples from $\boldsymbol{\theta}_{\#211}$ (\Cref{fig:simulation_for_two_cars}). In the comparison (\Cref{fig:ts_diag}), our method shows tighter containment of ground truth within the envelope of posterior motion states, while MA-IDM requires a wider range.
Quantitatively, we evaluate the model performance by comparing the absolute root mean square errors (RMSE) on motion states (acceleration, speed, and gap) between ground truth and fully simulated trajectories. Additionally, we use the continuously ranked probability score (CRPS) \citep{matheson1976scoring} to evaluate the performances of stochastic simulations and quantify the uncertainty of posterior motion states, which can be written as
\begin{equation}\label{crps}
    \mathrm{CRPS}(y_t) = \int_{-\infty}^{+\infty}\left(F(y)-\mathbbm{1}\{y>y_t\}\right)^2 dy,
\end{equation}
where $y_t$ is the observation at time $t$, $F$ is the forecast cumulative distribution function, and $\mathbbm{1}$ is the indicator function. To evaluate the effectiveness of the short-term simulation, we conducted simulations over all of the $20$ vehicles and use RMSE/CRPS of many fraction samples with 5 s, as shown in \Cref{tab:errors}. Results indicate that our model outperforms Bayesian IDM and MA-IDM in 5-second simulations, especially with the AR order $p\geq4$. In what follows, we mainly demonstrate the simulation performance using the model with AR(5). Recall that the length scale of the SE kernel in MA-IDM is $\ell \approx 1.44$ s \citep{zhang2024bayesian}, implying that MA-IDM generally performs well within $3\ell$ simulations. To evaluate this, we compare the simulations performances of the Bayesian IDM, the MA-IDM, and the Dynamic IDM ($p=5$) with fraction length varying from 1 s to 10 s, as shown in \Cref{fig:1to10_sim}. The results indicate that the MA-IDM could perform reasonably well within 4 s. But for those longer than $4$ s, the simulations with MA-IDM would then be more dominated by random noise rather than IDM, and its performance tends to be similar to the Bayesian IDM.

\begin{figure}[t]
    \centering
    \subfigure[Simulation with fixed IDM parameters and random white noise.]{
        \centering  \includegraphics[width=.65\textwidth]{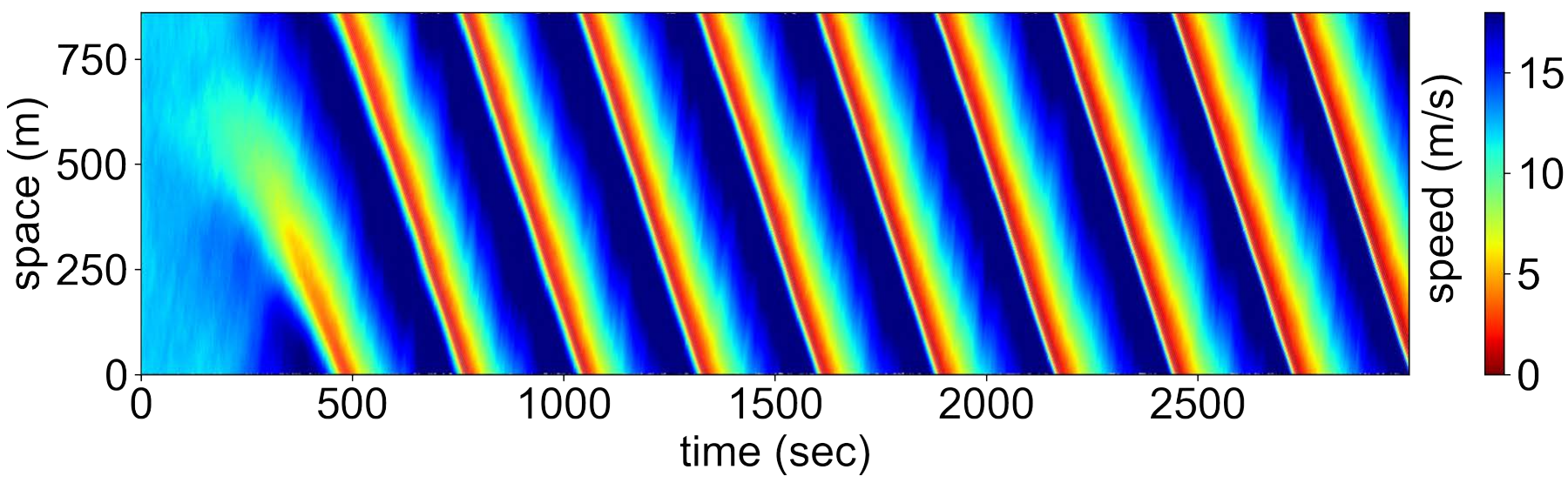}
    }\\
    \subfigure[Light traffic simulation with dynamic IDM ($p=5$).]{
        \centering\includegraphics[width=.65\textwidth]{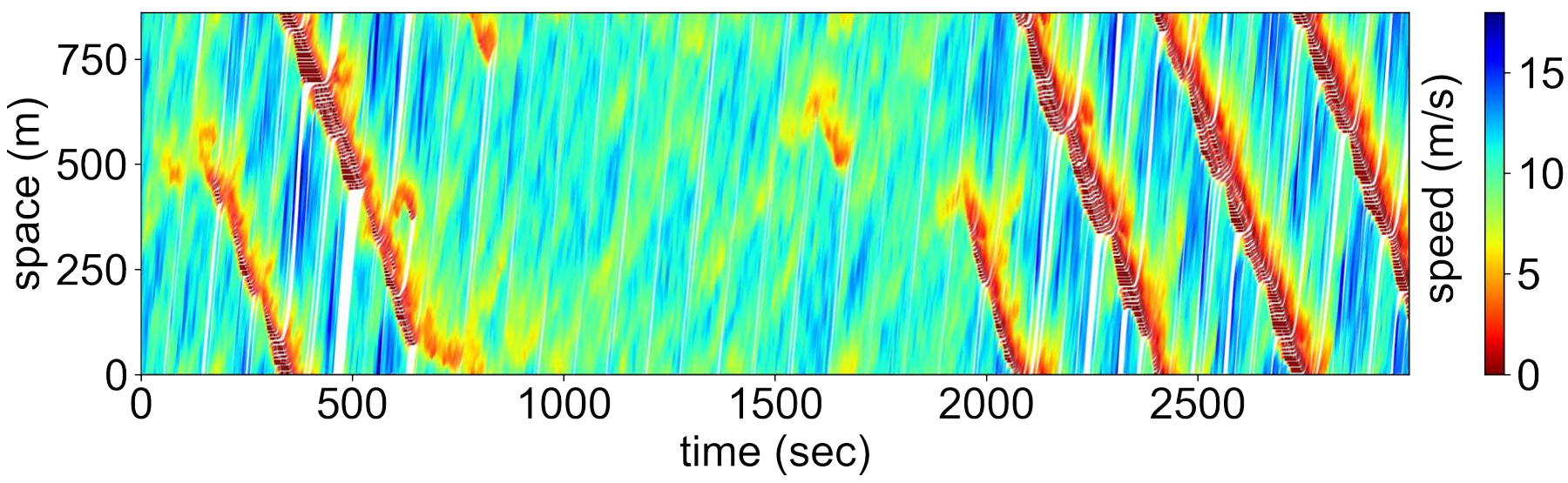}
    }\\
    \subfigure[Dense traffic simulation with dynamic IDM ($p=5$).]{
        \centering\includegraphics[width=.65\textwidth]{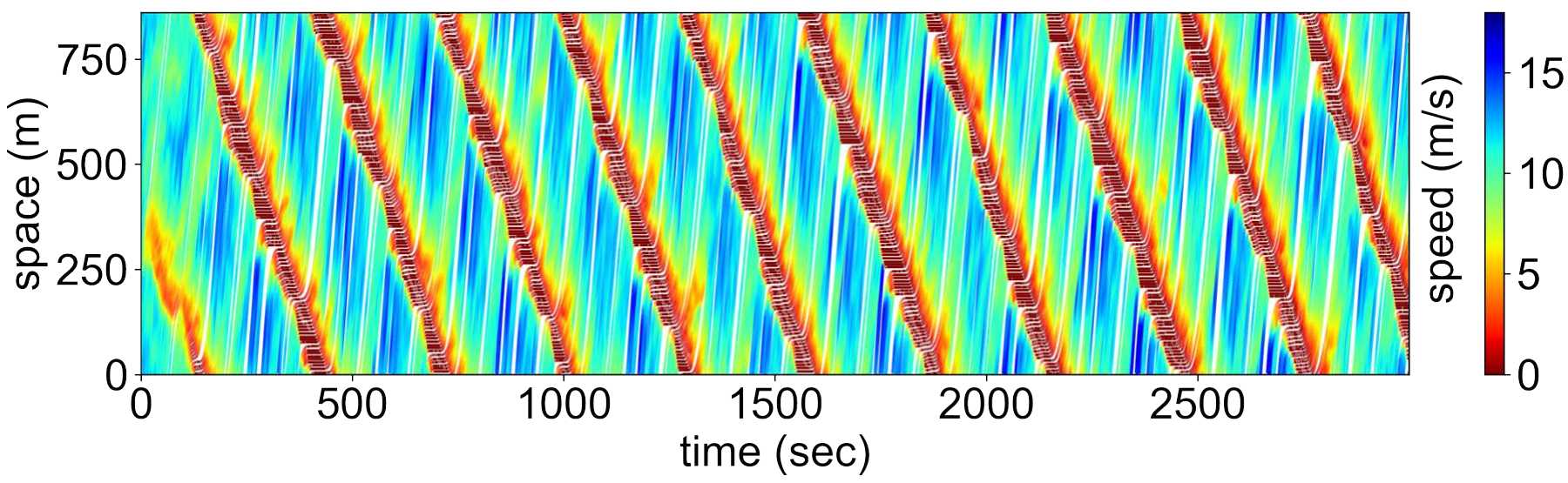}
    }
    \caption{Time-space diagram of multi-vehicle car-following simulations. The parameters are sampled from the posteriors of (a) the hierarchical Bayesian IDM ($p=0$) and (b) the hierarchical dynamic IDM ($p=5$). (c) Simulation with dense traffic.}
    \label{fig:multi_vehicle_simulation}
\end{figure}

\begin{figure}[ht]
    \centering
    \includegraphics[width = \linewidth]{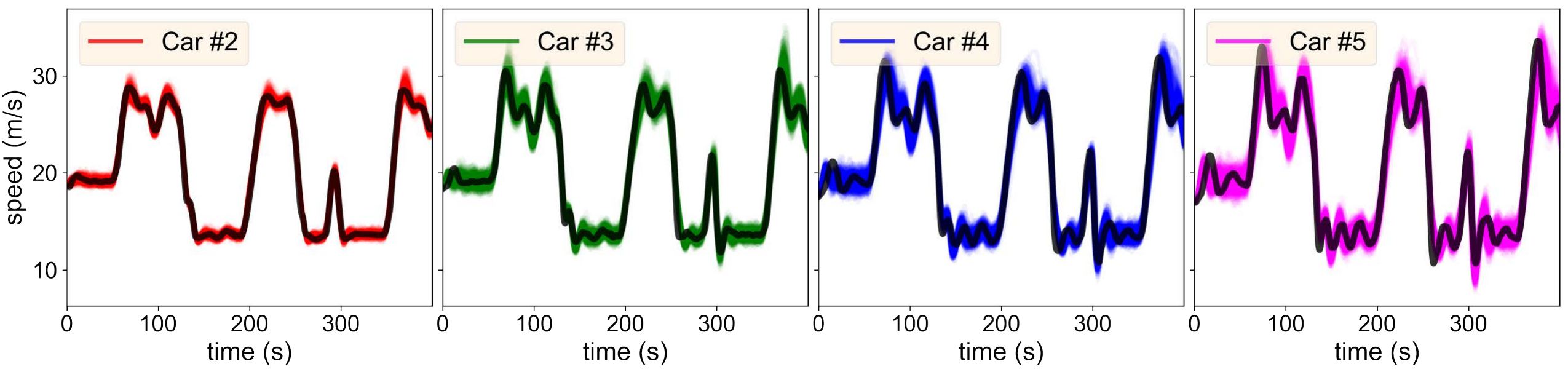}
    \caption{The multi-vehicle simulation in a platoon. Each vehicle's initial state is set to be the same as it was at 240 seconds (refer to the shaded part in \Cref{openacc_data}) in the OpenACC dataset.}
    \label{Platoon}
\end{figure}

\subsubsection{Multi-Vehicle Simulations: Ring-Road and Platoon}
Long-term simulations enable the analysis of dynamic traffic behaviors at the macroscopic level by simultaneously simulating a group of vehicles. To validate the capability for large-scale traffic simulations, we conduct multi-vehicle simulation experiments in a ring-road scenario \citep{sugiyama2008traffic} and a platoon.

\paragraph{Ring Road} Key elements of the ring road include a ring radius of $128$ m, resulting in a circumference of approximately $804$ m, initial speeds set at $11.6$ m/s, $32$ vehicles for light traffic and $37$ vehicles for dense traffic simulated for $15000$ steps with simulation step as $0.2$ s. The multi-vehicle simulation is conducted with two different models, as shown in \Cref{fig:multi_vehicle_simulation}. The top is simulated with Bayesian IDM and the others are simulated with the dynamic IDM ($p=5$). \Cref{fig:multi_vehicle_simulation} (a) presents a recurring pattern from the simulations with IDM parameters, although a stochastic term is introduced. On the contrary, different random car-following behaviors in the heterogeneous setting (see \Cref{fig:multi_vehicle_simulation} (b)) are obtained, indicating the drivers' dynamic and diverse driving styles. Dynamic IDM simulation can produce various traffic phenomena in the real world, including shock wave dissipation.

\paragraph{Platoon} In the platoon simulation, the first vehicle is the leader with its entire trajectory available. We simultaneously simulate the platoon's (i.e., four successive followers') trajectories based on the dynamic IDM with AR(5). \Cref{Platoon} shows the four followers' speed profiles, where the black lines represent the actual driving speed. We can see that the dynamic IDM still can accurately capture and replicate the real car-following behaviors even in long-range simulations.

\section{Conclusion}\label{conclusion}
This paper presents a dynamic regression framework for calibrating and simulating car-following models, which plays a critical role in understanding traffic flow dynamics. Our approach addresses a key limitation of existing car-following models by incorporating historical information, resulting in a more accurate reproduction of real-world driving behavior. By integrating AR processes into the error modeling, we offer a statistically rigorous alternative to the assumption of independent errors commonly found in current models. This enables the consideration of higher-order historical information, leading to improved simulation and prediction accuracy. The integration of AR processes into the Bayesian calibration framework is not merely an addition but a significant enhancement that addresses specific limitations of the previous models. This integration allows for a more nuanced and dynamic understanding of time-dependent stochastic residuals (especially capturing the negative correlations), which is a substantial step forward in the calibration of car-following models.

Experiment results indicate that modeling human car-following behavior should incorporate actions from the past 10 s, capturing short-term positive correlations ($0\sim5$ s) and long-term negative correlations ($5\sim10$ s) in real driving data. The framework's effectiveness is demonstrated through its application to the HighD data. The proposed framework preserves the parsimonious nature of traditional car-following models while offering enhanced probabilistic simulations. Although we used IDM as an example to illustrate our idea, it is worth noting that this idea is flexible to extend to calibrating many other traditional car-following models, this may potentially improve the performance while without modifying their model structures. To sum up, our research sheds light on a potential direction for the development of high-fidelity microscopic traffic simulation models. By integrating historical data, our dynamic regression framework enables more accurate predictions and simulations of car-following behavior, which is crucial for traffic management and planning. In addition, this work contributes to the broader field of traffic theory and simulation by providing a more robust and accurate tool (which is open-sourced) for researchers and practitioners. We hope our findings could stimulate further research in this direction, fostering a deep understanding of traffic flow dynamics and the development of advanced and efficient microscopic traffic models.

However, as with all research, there are avenues for further development. Future work could explore incorporating a mean model as a car-following model with time-varying parameters to enhance adaptability to diverse scenarios and driving modes. Additionally, adding a moving average component might capture the temporal dependence and irregular patterns, and reduce the noise level through smoothing.

\section*{CRediT authorship contribution statement}
\textbf{Chengyuan Zhang}: Conceptualization, Methodology, Formal Analysis, Writing - Original Draft,
Writing - Review \& Editing. \textbf{Wenshuo Wang}: Conceptualization, Methodology, Formal Analysis, Writing - Original Draft, Writing - Review \& Editing. \textbf{Lijun Sun}: Conceptualization, Methodology, Formal Analysis, Writing - Original Draft, Writing - Review \& Editing, Supervision, Funding acquisition.

\section*{Declaration of competing interest}
The authors declare that they have no known competing financial interests or personal relationships that could have appeared to influence the work reported in this paper.

\section*{Acknowledgement}
This research is supported by the Natural Sciences and Engineering Research Council of Canada (NSERC) of Canada (Discovery Grant RGPIN-2019-05950). C. Zhang would like to thank the McGill Engineering Doctoral Awards (MEDA), the Interuniversity Research Centre on Enterprise Networks, Logistics and Transportation (CIRRELT), Fonds de recherche du Québec -- Nature et technologies (FRQNT), and NSERC for providing scholarships and funding to support this study.

\section*{Appendix}
\subsection*{Details for the calibration methods}
\subsubsection*{(1) Calibration on speed data with observation noise}
Recall that our model follows
\begin{align}
    a^{(t)} =& \, \mathrm{IDM}^{(t)}+ \epsilon^{(t)}, \nonumber\\
    \epsilon^{(t)} =& \, \rho_{d,1}\epsilon^{(t-1)} + \rho_{d,2}\epsilon^{(t-2)}+\dots + \rho_{d,p}\epsilon^{(t-p)} + \eta^{(t)}, \nonumber\\
    \eta^{(t)} \,\stackon{$\sim$}{\tiny{\textit{i.i.d.}}} \,& \mathcal{N}(0,\sigma^2_\eta), \nonumber
\end{align}
and given the dynamic updating in \Cref{update_v} ${v}^{(t+1)} = v^{(t)}+ a^{(t)} \Delta t$, then we have
\begin{subequations}
\begin{align}
    a^{(t)} =& \, \mathrm{IDM}^{(t)}+ \rho_{d,1}\left(a^{(t-1)} - \mathrm{IDM}^{(t-1)}\right) + \rho_{d,2}\left(a^{(t-2)} - \mathrm{IDM}^{(t-2)}\right)\nonumber\\ &\,+
    \cdots+ \rho_{d,p}\left(a^{(t-p)} - \mathrm{IDM}^{(t-p)}\right) +\eta^{(t)},\\
    v^{(t+1)} =& \,v^{(t)}+ \mathrm{IDM}^{(t)}\Delta t+ \rho_{d,1}\left(a^{(t-1)}\Delta t - \mathrm{IDM}^{(t-1)}\Delta t\right) + \cdots \nonumber\\ &\,+ \rho_{d,p}\left(a^{(t-p)}\Delta t - \mathrm{IDM}^{(t-p)}\Delta t\right) +\eta^{(t)}\Delta t \\
    =&\,v^{(t)}+ \mathrm{IDM}^{(t)}\Delta t+ \rho_{d,1}\left((v^{(t-1)} + a^{(t-1)}\Delta t) - (v^{(t-1)} + \mathrm{IDM}^{(t-1)}\Delta t)\right)+
    \cdots \nonumber\\
    &\, +\rho_{d,p}\left((v^{(t-p)} + a^{(t-p)}\Delta t) - (v^{(t-p)} + \mathrm{IDM}^{(t-p)}\Delta t)\right)  +\eta^{(t)}\Delta t \\
    =&\, v^{(t)}+ \mathrm{IDM}^{(t)}\Delta t+ \rho_{d,1}\left(v^{(t)} - (v^{(t-1)} + \mathrm{IDM}^{(t-1)}\Delta t)\right) +
    \cdots \nonumber\\
    &\, + \rho_{d,p}\left(v^{(t-(p-1))} - (v^{(t-p)} + \mathrm{IDM}^{(t-p)}\Delta t)\right) +\eta^{(t)}\Delta t\\
    =&\, v^{(t)}+ \mathrm{IDM}^{(t)}\Delta t+ \sum_{k=1}^p \rho_{d,k} \left(v^{(t-(k-1))} - v^{(t-k)}\right) - \sum_{k=1}^p \rho_{d,k} \mathrm{IDM}^{(t-k)}\Delta t +\eta^{(t)}\Delta t.\label{calibrate_v_results}
\end{align}
\end{subequations}
Such that we can give the likelihood of a noisy observation as
\begin{equation}
    \hat{v}^{(t+1)}\sim \mathcal{N}\left(\bar{v}^{(t+1)}, (\sigma_\eta\Delta t)^2+\sigma_v^2\right),
\end{equation}
where the mean
\begin{equation}
    \bar{v}^{(t+1)}=v^{(t)}+ \mathrm{IDM}^{(t)}\Delta t+ \sum_{k=1}^p \rho_{d,k} \left(v^{(t-(k-1))} - v^{(t-k)}\right) - \sum_{k=1}^p \rho_{d,k} \mathrm{IDM}^{(t-k)}\Delta t,\nonumber
\end{equation}
$\hat{v}^{(t+1)}$ is the observed data of the true speed ${v}^{(t+1)}$, and $\sigma_v^2$ is the variance of the observation noise.

\subsubsection*{(2) Calibration on position/spacing data with observation noise}
Similar to the previous approach, from \Cref{update_s} we have $x^{(t+1)} = x^{(t)} + \frac{1}{2}\left(v^{(t)}+v^{(t+1)}\right)\Delta t$; then according to \Cref{calibrate_v_results}, we can reformat $x^{(t+1)}$ as
\begin{subequations}
\begin{align}
    x^{(t+1)} =&\,  x^{(t)} + \frac{1}{2}v^{(t)}\Delta t + \frac{1}{2}\biggl(v^{(t)}+ \mathrm{IDM}^{(t)}\Delta t+ \sum_{k=1}^p \rho_{d,k} \left(v^{(t-(k-1))} - v^{(t-k)}\right) \\ &  - \sum_{k=1}^p \rho_{d,k} \mathrm{IDM}^{(t-k)}\Delta t +\eta^{(t)}\Delta t\biggr)\Delta t \nonumber\\
    =&\, x^{(t)} + v^{(t)}\Delta t + \frac{1}{2}\mathrm{IDM}^{(t)}\Delta t^2 + \frac{1}{2}\sum_{k=1}^p \rho_{d,k} \left(v^{(t-(k-1))} - v^{(t-k)}\right)\Delta t \\ &  - \frac{1}{2}\sum_{k=1}^p \rho_{d,k} \mathrm{IDM}^{(t-k)}\Delta t^2 +\frac{1}{2}\eta^{(t)}\Delta t^2. \nonumber
\end{align}
\end{subequations}
Accordingly, the likelihood is written as
\begin{equation}
    \hat{x}^{(t+1)} \sim \mathcal{N}\left(\bar{x}^{(t+1)}, \Bigl(\frac{1}{2}\sigma_\eta\Delta t^2\Bigr)^2+\sigma_x^2\right),
\end{equation}
where the mean
\begin{equation}
    \bar{x}^{(t+1)}=x^{(t)} + v^{(t)}\Delta t + \frac{1}{2}\mathrm{IDM}^{(t)}\Delta t^2 + \frac{1}{2}\sum_{k=1}^p \rho_{d,k} \left(v^{(t-(k-1))} - v^{(t-k)}\right)\Delta t  - \frac{1}{2}\sum_{k=1}^p \rho_{d,k} \mathrm{IDM}^{(t-k)}\Delta t^2,\nonumber
\end{equation}
$\hat{x}^{(t+1)}$ is the observed data of the true position ${x}^{(t+1)}$, and $\sigma_x^2$ is the variance of the observation noise of position data. Note that this is a position-based form, but one can easily adapt it into a gap-based form.

\vskip 0.2in
\bibliography{references}

\end{document}

%% file: Figs/decision_scheme.tex
\tikzset{every picture/.style={line width=0.75pt}} 

\begin{tikzpicture}[x=0.75pt,y=0.75pt,yscale=-1,xscale=1]

\node [circle, line width=.5mm, inner sep=0pt,minimum size=1.2cm] (t2) at (-60,-80) {$t-1$};
\node [circle, line width=.5mm, inner sep=0pt,minimum size=1.2cm] (t2) at (50,-80) {\color{gray} $t$};
\node [circle, line width=.5mm, inner sep=0pt,minimum size=1.2cm] (t2) at (160,-80) {\color{lightgray} $t+1$};

\node [circle, draw=gray, fill=black!10, line width=.5mm, inner sep=0pt,minimum size=1.2cm] (v2) at (55,30) {${v}^{(t)}$};

\node [circle, draw=gray, fill=black!10,line width=.5mm, inner sep=0pt,minimum size=1.2cm] (x2) at (50,-40) {${x}^{(t)}$};

\node [circle, draw=lightgray!70, fill=black!3,line width=.5mm, inner sep=0pt,minimum size=1.2cm] (a2) at (50,100) {$a^{(t)}$};

\node [circle, draw=black, fill=black!37,line width=.5mm, inner sep=0pt,minimum size=1.2cm] (v1) at (-55,30) {${v}^{(t-1)}$};

\node [circle, draw=black, fill=black!37,line width=.5mm, inner sep=0pt,minimum size=1.2cm] (x1) at (-60,-40) {${x}^{(t-1)}$};

\node [circle, draw=gray, fill=black!10,line width=.5mm, inner sep=0pt,minimum size=1.2cm] (a1) at (-60,100) {$a^{(t-1)}$};

\node [circle, draw=lightgray!70, fill=black!3,line width=.5mm, inner sep=0pt,minimum size=1.2cm] (v3) at (165,30) {${v}^{(t+1)}$};

\node [circle, draw=lightgray!70, fill=black!3,line width=.5mm, inner sep=0pt,minimum size=1.2cm] (x3) at (160,-40) {${x}^{(t+1)}$};

\node [circle, draw=lightgray!30,line width=.5mm, inner sep=0pt,minimum size=1.2cm] (a3) at (160,100) {$a^{(t+1)}$};

\node[mark size=1pt,color=black] at (-120,30) {$\cdots$};
\node[mark size=1pt,color=black] at (-120,-40) {$\cdots$};
\node[mark size=1pt,color=black] at (-120,100) {$\cdots$};
\node[mark size=1pt,color=black] at (220,30) {$\cdots$};
\node[mark size=1pt,color=black] at (220,-40) {$\cdots$};
\node[mark size=1pt,color=black] at (220,100) {$\cdots$};

\path [draw,-,color=blue] (x1) edge [bend right] node {} (a1.35);
\path [draw,->,color=blue] (v1) edge [bend right] node {} (a1.37);
\path [draw,->,color=red] (a1) edge [bend left] node {} (v2);
\path [draw,->,color=red] (a1) edge [bend left] node {} (x2.-135);
\path [draw,-,color=blue] (x2) edge [bend right] node {} (a2.35);
\path [draw,->,color=blue] (v2) edge [bend right] node {} (a2.37);
\path [draw,->,color=red] (a2) edge [bend left] node {} (v3);
\path [draw,->,color=red] (a2) edge [bend left] node {} (x3.-135);
\path [draw,-,color=blue] (x3) edge [bend right] node {} (a3.35);
\path [draw,->,color=blue] (v3) edge [bend right] node {} (a3.37);

\path [draw,->,color=black!37] (x1) edge node {} (x2);
\path [draw,->,color=black!37] (v1) edge node {} (v2);
\path [draw,->,color=black!37] (v1) edge node {} (x2);
\path [draw,->,color=black!37] (x2) edge node {} (x3);
\path [draw,->,color=black!37] (v2) edge node {} (v3);
\path [draw,->,color=black!37] (v2) edge node {} (x3);

\node [text width=3.cm] (t) at (315,-20) {\color{red} Action};
\path [draw,->,color=red] (325,-20) to (350,-20);
\node [text width=3.cm] (t) at (310,20) {\color{blue} Decision};
\path [draw,->,color=blue] (325,20) to (350,20);
\end{tikzpicture}